\documentclass[twocolumn,preprintnumbers,amsmath,amssymb]{revtex4-1}

\usepackage{graphicx}
\usepackage[breaklinks=true,colorlinks=true,linkcolor=blue,urlcolor=blue,
citecolor=blue]{hyperref}

\begin{document}

\newcommand{\sio}{SiO$_2$}
\newcommand{\water}{H$_2$O}
\newcommand{\watero}{H$_2$O$_2$}
\newcommand{\slp}{SLP}


\title{H$_{2}$O incorporation in the phosphorene/a-SiO$_{2}$ interface:\\A
first-principles study}

\author{Wanderl\~a L. Scopel}
\email{wlscopel@if.uff.br}
\affiliation{Departamento de F\'{\i}sica, Universidade Federal do Esp\'irito 
Santo, Vit\'oria, ES, 29075-910, Brazil}
\affiliation{Departamento de Ci\^encias Exatas, Universidade Federal 
Fluminense, 
Volta Redonda, RJ, 27255-250, Brazil}

\author{Everson S. Souza}
\email{nosreveazuos@gmail.com}
\affiliation{Departamento de F\'{\i}sica, Universidade Federal do Esp\'irito 
Santo, Vit\'oria, ES, 29075-910, Brazil}

\author{R. H. Miwa}
\email{hiroki@infis.ufu.br}
\affiliation{Departamento de F\'{\i}sica, Universidade Federal de Uberl\^andia, 
Caixa Postal 593, CEP 38400-902, Uberl\^andia, MG, Brazil}

\date{\today}

\begin{abstract}

Based on first-principles calculations, we investigate the energetic stability 
and the electronic properties of (i) a single layer phosphorene (SLP) adsorbed 
on the amorphous \sio\ surface (\slp/a-\sio), and (ii) the further 
incorporation of water molecules at the phosphorene/a-\sio\ interface. In (i), 
we find that the phosphorene sheet bonds to a-\sio\  through  van der Waals 
interactions, even upon the presence of oxygen vacancy on the surface. The 
\slp/a-\sio\ system presents a type-I band alignment, with the valence 
(conduction) band maximum (minimum) of the phosphorene lying within the energy 
gap of the a-\sio\ substrate. The structural, and the surface-potential 
corrugations promote  the formation of electron-rich and -poor regions on the 
phosphorene sheet and at the \slp/a-\sio\ interface. Such charge density puddles 
have been strengthened by the presence of oxygen vacancies in a-\sio. In (ii), 
due to the amorphous structure  of the surface, we have considered a number of 
plausible geometries of \water\ embedded in the \slp/a-\sio\ interface. There is 
an energetic preference to  the formation of hydroxyl (OH) groups on the a-\sio\ 
surface. Meanwhile, upon the presence of oxygenated water or interstitial oxygen 
in the phosphorene sheet, we find the formation of metastable OH bonded to the 
phosphorene, and the formation of energetically stable P--O--Si chemical bonds 
at the \slp/a-\sio\ interface. Further x-ray absorption spectra (XAS) 
simulations have been done, aiming to provide additional structural/electronic 
informations of the  oxygen atoms forming hydroxyl groups or P--O--Si chemical 
bonds at the interface region.

\end{abstract}
 


\maketitle

\section{Introduction}

Phosphorene is a two-dimensional (2D) material,  consisting of a single-layer of 
the black phosphorus allotrope that, currently,  has received tremendous 
attention. It presents unique properties, such as layer-controlled direct 
bandgap\,\cite{zhang2014extraordinary,tran2014layer}, 
high-mobility\,\cite{qiao2014high}, and \textit{p}-type semiconductor 
properties\,\cite{liu2014phosphorene}. These  properties make  phosphorene a quite 
interesting 2D material  for electronic 
applications\,\cite{li2014black,das2014ambipolar,zhu2015flexible,du2014device, 
du2015recent}.

There are two main methods  to produce phosphorene, both based on the 
exfoliation processes down to a monolayer thickness,  {\it viz.}: 
mechanical exfoliation\,\cite{verzhbitskiy2015colossal,favron2015photooxidation, 
lu2014plasma} and liquid exfoliation\,\cite{guo2015black, 
brent2014production,kang2015solvent}. However, exfoliated phosphorus degrades 
when exposed in ambient 
conditions\,\cite{koenig2014electric,woomer2015phosphorene}. Such a phosphorene 
degradation\,\cite{favron2015photooxidation,island2015environmental} is an 
irreversible process,  ascribed to the presence of oxygen atoms, and water 
molecules. Indeed, phosphorene presents lower chemical 
stability due to the lone pairs of the surface P atoms, which leads it to be 
very reactive when exposed to air. In Ref.\,\cite{ziletti2015oxygen} the authors 
have been shown that oxygen 
interstitial defect on phosphorene lead to increasing of hydrophilicity of the 
surface, which may  play an important role in the surface degradation. Further 
studies have been done in an  effort to understand the degradation of 
phosphorene upon the presence of O$_2$ and/or H$_2$O 
molecules\,\cite{cai2015energetics,favron2015photooxidation,wang2016degradation, 
wood2014effective}, as well as some proposals to avoid it. For instance, 
phosphorene surface passivation through Al$_2$O$_3$ 
coating\,\cite{na2014few,kim2014toward,pei2016producing}.

The degradation process of phosphorene is strongly correlated with its boundary 
conditions. In a recent experimental study, Wood {\it et 
al.}\,\cite{wood2014effective} verified that the degradation of transistors, 
composed by few layers of phosphorene,  can be suppressed upon the presence of a 
hydrophilic substrate (\sio) and a coating layer of AlO$_x$. Since \sio\ surface 
is by far the most common substrate to buildup transistors, it is quite 
important to get a clear  picture of the electronic interactions and the 
structural properties of (i) phosphorene adsorbed on the \sio\ surface, as well 
as (ii) the role played the \sio\ surface in the phosphorene degradation 
mediated by the presence of \water\ molecules.

In this work, we performed an atomistic study, based on first-principles 
calculations of  (i) and (ii) described above. We have considered a single 
layer phosphorene (\slp) adsorbed on the amorphous \sio\ surface (\slp/a-\sio). 
In (i), we found that the \slp\ bonds to the a-\sio\ surface mediated by van 
der Walls (vdW) interactions, even upon the presence of oxygen vacancies ($\rm 
V_O$), which is the most common intrinsic defect in \sio. The structural, and 
the surface-potential corrugations promote  the formation of electron-rich and 
-poor regions on the phosphorene sheet, and at the phosphorene/a-\sio\ 
interface. Such charge density puddles are strengthened by the presence of 
oxygen vacancies  in a-\sio. In (ii), we have considered a number of plausible 
configurations for the \water\ molecules embedded in the \slp/a-\sio\ interface. 
We find an energetic preference for the formation of hydroxyl (OH) groups bonded 
to the a-\sio\ surface, when compared with the \slp,  even upon the presence of 
the interstitial oxygen adatoms in the \slp\ sheet. However, due to the 
amorphous character of the a-\sio\ surface, depending on the local geometry at 
the \slp/a-\sio\ interface region, we verified the formation of energetically 
stable  P--O--Si structure bridging the \slp\ sheet and the a-\sio\ surface.
Finally,  x-ray absorption spectra (XAS) 
simulations have been done, aiming to provide additional structural/electronic 
informations of the  oxygen atoms forming hydroxyl groups or P--O--Si chemical 
bonds at the interface region.

\section{Methodology}

The amorphous structure was generated through \textit{ab initio} molecular 
dynamics (MD) simulations based on the density functional theory (DFT) approach, 
as implemented in the VASP code\cite{kresse1,kresse2,kresse3}. In Ref. 
\cite{scopelPRB2008}, we present details on the generation procedure of 
amorphous SiO$_2$ bulk structure. For DFT calculatons, the generalized gradient 
approximation (GGA) for the exchange-correlation potential is used. The 
ion-electron interaction is treated with the projected augmented wave 
(PAW)\cite{bl1994p,kresse1999ultrasoft} method. The plane-wave cutoff energy for 
wave function is set to 500 eV and the brillouin zone was sampled at the 
$\Gamma$ point. All atoms were allowed to relax until the atomic forces were 
smaller than 0.025 eV/\AA. In addition, a combination of optB88vdW 
\cite{dion2004van,klimevs2011van} for geometry optimization and HSE06 
\cite{ernzerbhof2006} for density of state (based on the optB88-vdW optimized 
structure) is used, which has been shown very reliable for single-layer 
phosphorene (SLP). The slab model contains three different atomic species with 
vacuum region of ca. 15 \AA~, so that the interaction between successive 
periodic images along to z-direction can be neglected. The optimized lattice 
parameter for single-layer phosphorene (SLP) was a = 4.38 \AA~ and b = 3.31 \AA, 
in good agreement with previous theoretical and experimental 
results\cite{liu2014phosphorene,brown1965refinement}. In our calculations we 
used a 3$\times$4 supercell size.

\section{Results and Discussions}

\subsection{SLP/a-SiO$_2$ interface}

\begin{figure}[!ht]
\begin{center}
\includegraphics[width = 8.5 cm]{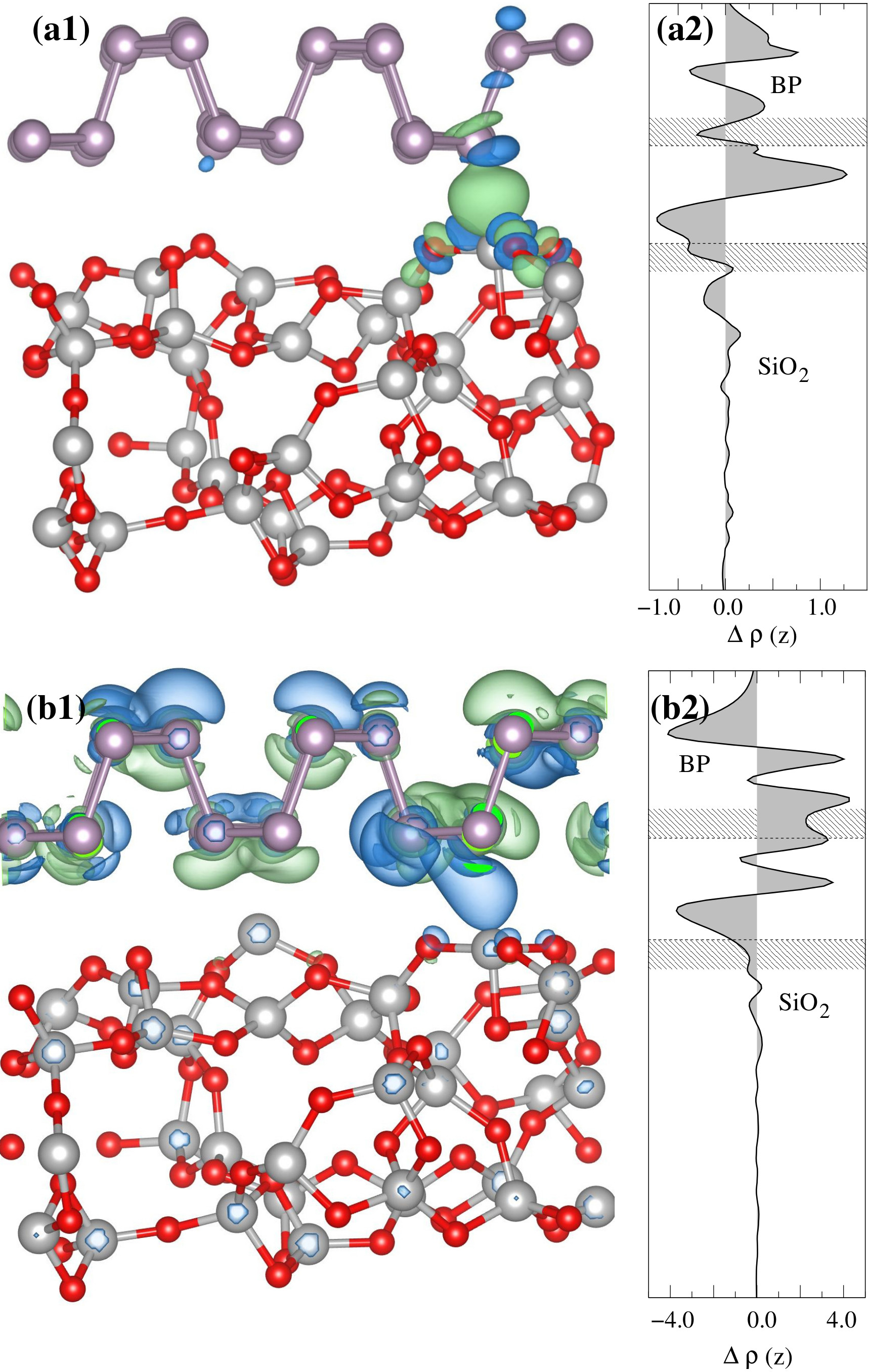}
\caption{(a1) Structural model and the electronic charge transfers 
($\Delta\rho$) of a single layer phosphorene adsorbed on the a-\sio\ surface 
(\slp/a-\sio); (a2) planar average of the net charge transfer perpendicularly to 
the \slp/a-\sio\ interface. (b1) Structural model and the electronic charge 
tranfers of \slp/a-\sio\ upon the presence of an intrinsic defect on the 
surface, oxygen vacancy; (b2)   planar average of the net charge transfer 
perpendicularly to the \slp/a-\sio\ interface. Green regions indicate an 
increase of total charge density, $\Delta \rho >$ 0, while blue regions indicate 
a decrease ($\Delta \rho <$ 0). The P, O, Si and H atoms are represented by 
light purple, red and gray color spheres. }
\label{structure}
\end{center}
\end{figure}

Initially we examine the  energetic stability of a \slp\ adsorbed on the 
defect-free amorphous \sio\ surface (\slp/a-\sio). In Fig.\,\ref{structure}(a1) 
we present the structural model and the electronic charge transfers (which will 
be discussed below) of \slp/a-\sio. The energetic stability of the SLP adsorbed 
on the SiO$_2$ surface was examined by the calculation of the adsorption energy 
($E^{\rm a}$), which can be written as,

$$
E^{\rm a} =  E[{\rm a\textendash SiO_2}] + E[{\rm SLP}] - E[{\rm SLP/a 
\textendash SiO_2}].
$$

Where $E[{\rm SLP}]$ and $E[{\rm a\textendash SiO_2}]$ are  the total energies 
of the separated components, an isolated \slp, and the a-\sio\ surface, 
respectively; $E[{\rm SLP/a\textendash SiO_2}]$  is the total energy of the 
fully relaxed SLP/a-SiO$_2$ system. It was considered the adsorption of SLP on 
two different (amorphous) \sio\ surfaces, where we found adsorption energies of 
13.9 and 11.3\,eV/\AA$^2$, and (averaged) vertical distances between the \slp\ 
and the a-\sio\ surface of 2.80 and 2.75\,\AA. Thus, suggesting the absence of  
chemical bonds at the \slp--surface region. The \slp\ sheet is attached to 
the a-\sio\ surface mediated by  vdW interactions. Somewhat similar picture has 
been verified for other 2D systems adsorbed on a-\sio. For instance, we obtained 
$E^{\rm a}$ of 6.3 and 15\,meV/\AA$^2$ for graphene and MoS$_2$ on the a-\sio\ 
surface,\,\cite{miwaAPL2011,scopel2015mos2}.

Due to the surface corrugation, \slp\ on a-\sio\ may also present structural 
deformations, as observed  for graphene on 
a-\sio\,\cite{ishigamiNanoLett2007,sinitskiiACSNano2010}; giving rise to 
electron-rich and -poor regions (so called electron-hole puddles) on the 
graphene surface\,\cite{martinNatPhys2008}. Indeed, due to corrugation of the 
(amorphous) surface potential, and the vertical distortion of $\sim$0.08\,\AA\ 
of the \slp\ adsorbed on a-\sio, we also verify the formation of electron-hole 
puddles on the \slp\ surface, as well as electronic charge transfers at the   
\slp$-$a-\sio\ interface region. Here, we map  the total charge transfers 
($\Delta\rho$) by comparing total charge density of the final system, 
\slp/a-\sio\ ($\rm\rho$[SLP/a\hbox{-}SiO$_2$]) with the ones of the isolated 
components,  SLP ($\rm\rho[SLP]$) and a-\sio\ ($\rm\rho[a\hbox{-}SiO_2]$),

$$
\rm 
\Delta\rho= \rho[SLP/a\hbox{-}SiO_2]-\rho[SLP]-\rho[a\hbox{-}SiO_2].
$$

Our result of $\Delta\rho$ for a \slp\ on the pristine a-\sio\ surface is 
presented in Fig.\,\ref{structure}(a1). In Fig.\,\ref{structure}(a2) we present 
the planar average of $\Delta\rho$ perpendicularly to the surface plane, 
$\Delta\rho(z)$. The electronic charge transfer is (i) not uniform on the 
surface plane, and (ii) mostly localized at the interface region, where we have 
both positive as negative values of $\Delta\rho$. In order to quantify the total 
charge transfers, we have used the  Bader charge density 
analysis\,\cite{bader1,bader2}; where we found that the total charge density of 
the \slp\ reduces by $6.70\times 10^{12} e/{\rm cm^2}$.

The \slp$-$\sio\ interaction has been strengthened upon the presence of  oxygen 
vacancies. We found that the adsorption energy increases to 16.6 meV/\AA${^2}$, 
and there is a  net charge transfer  of $1.06\times 10^{13}e/{\rm cm^2}$ 
(0.19\,$e$/$\rm V_O$ defect), from the surface to the \slp. In 
Figs.\,\ref{structure}(b1) and \ref{structure}(b2), we present our results of 
$\Delta\rho$ and $\Delta\rho(z)$, respectively. It noticeable that the most of 
the charge density gain is localized on the lower zigzag chains of the 
phosphorene; somewhat similar picture has been verified for the the n-type doped 
graphene bilayers on the Cu(111) surface\,\cite{souza2016switchable}. It is 
worth noting the formation electron-rich ($\Delta\rho>0$, green regions) and 
electron-poor ($\Delta\rho<0$, blue regions) embedded in the phosphorene layer, 
Fig.\,\ref{structure}(b1), while $\Delta\rho\approx 0$ in the \sio\ surface. As 
verified for graphene on \sio\,\cite{martinNatPhys2008,miwaAPL2011},  those 
electron-rich and -poor regions, randomly distributed on the phosphorene layer, 
are ruled by the structural deformation of the adsorbed \slp, and the 
corrugation of the a-\sio\ surface potential.

\begin{figure}[!ht]
\begin{center}
\includegraphics[width = 8.5 cm]{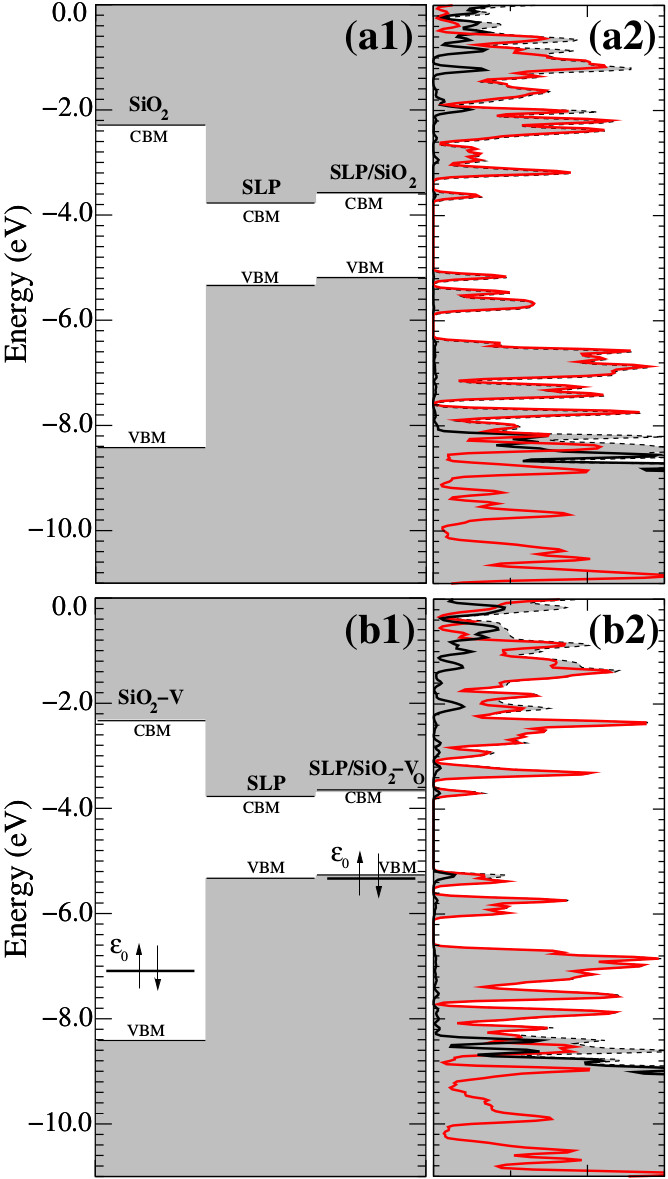}
\caption{Electronic energy levels of the valence band maximum (VBM) and the 
conduction band minimum (CBM) of the isolated components, \slp\ and a-\sio, and 
the ones of the final system, \slp/a-\sio, and the projected density of states 
(PDOS), of the pristine surface (a1-a2), and  upon the presence of an oxygen 
vacancy (b1-b2).  $\varepsilon_0$ indicate the energy position of the (oxygen 
vacacy) defect level.} 
\label{energies_level}
\end{center}
\end{figure}

Focusing on the electronic properties,  we calculate the energy positions of the 
valence band maximum  (VBM), and conduction band minimum (CBM) of the isolated 
components, and the ones of the final system, \slp/a-\sio, 
Fig.\,\ref{energies_level}. The energy positions were aligned with respect to 
the vacuum level\,\cite{vacuum-level}. For the \slp\ we find an ionization 
potential of 5.3\,eV, and an energy gap of 1.55\,eV, which are in good agreement 
with recent theoretical 
studies\,\cite{cai2014layer,srivastava2015tuning,tran2014layer}. Meanwhile, for 
the a-\sio\ we find an energy gap of 6.15\,eV, and an ionization potential of 
8.4\,eV. In Fig.\,\ref{energies_level}(a2) we present the density of states 
(DOS) (shaded region) and the projected density of states (PDOS) on the \slp\ 
(dashed lines) and \sio\ (solid lines). The VBM and the CBM of the \slp\ are 
slightly perturbed (by about 0.1\,meV) upon its interaction with the a-\sio\ 
surface, being both localized within the energy gap of the a-\sio. In this case, 
 the \slp$-$a-\sio\ interface presents a type-I band alignment, with  valence 
band offset (VBO) and conduction band offset (CBO) of about 3.11 and 1.47\,eV, 
respectively. Here we are not considering the contribution of the dipole 
effects, at the \slp$-$a-\sio\ interface region, on the band VBM and CBM band 
alignment\,\cite{dipole}.

A single oxygen vacancy on the a-\sio\ surface gives rise to  an occupied  
defect level within the energy gap of a-\sio. Here, we find the an occupied 
defect level lying at VBM+1.60\,eV [Fig.\,\ref{energies_level}(b1)], in good 
agreement  with  previous theoretical work\,\cite{sushko2005oxygen}. Upon the 
formation of \slp/a-\sio, the defect level is resonant with  the VBM of the adsorbed \slp\ sheet, 
Figs.\,\ref{energies_level}(b1) and \ref{energies_level}(b2); meanwhile the 
energy positions of the VBM and CBM of \slp\ are weakly perturbed by the 
presence of the $\rm V_O$ defect.

\subsection{H$_2$O at SLP/SiO$_2$ interface}

\begin{figure}[!ht]
\begin{center}
\includegraphics[width = 8.5 cm]{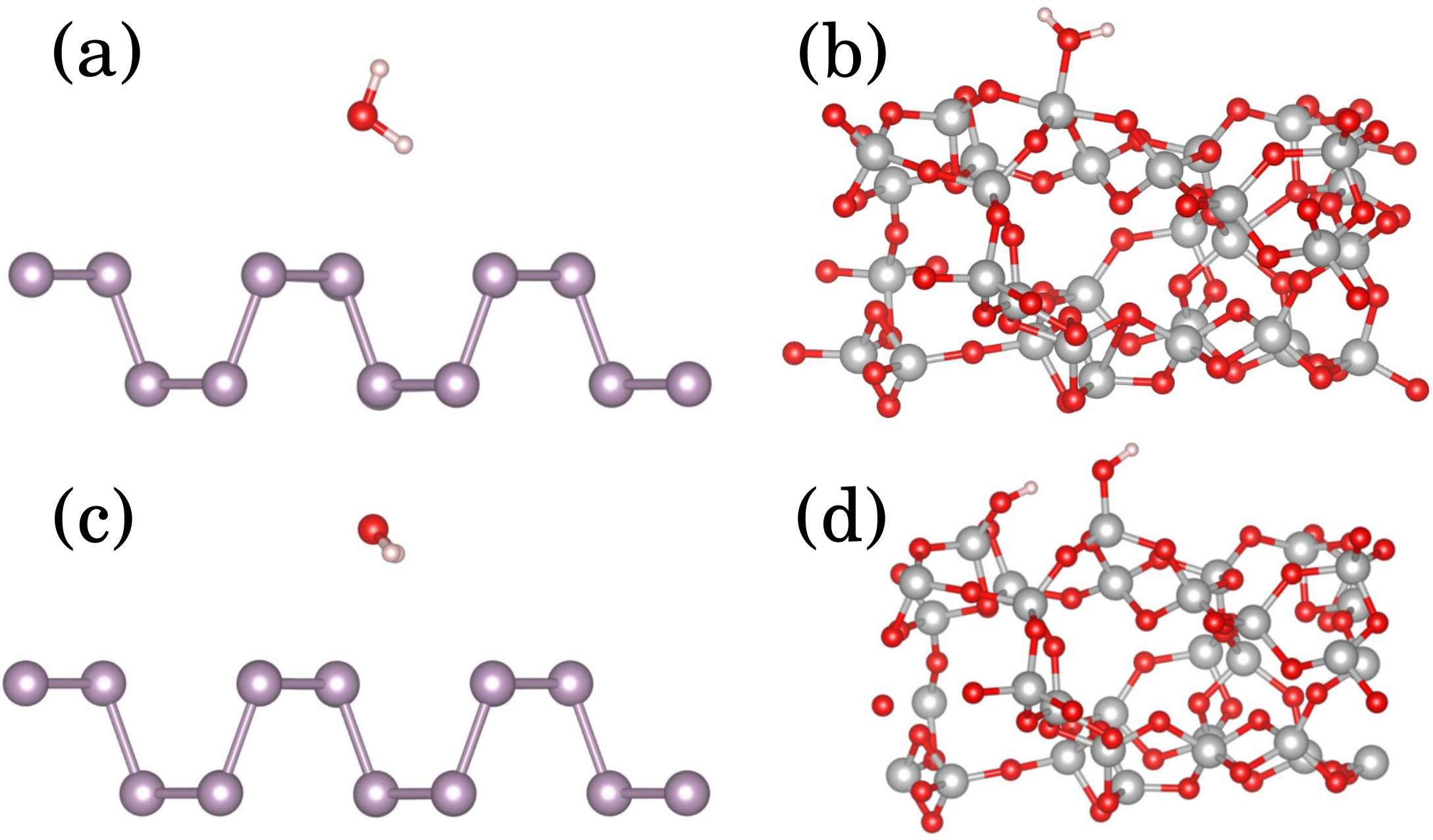}
\caption{Equilibrium  geometries of the H$_2$O molecule adsorbed on the  
single layer phosphorene (a) and (c); and on the a-\sio\ surface (b) and (d). 
The P, O, Si and H atoms are represented by light purple, red, gray and white 
color spheres.}
\label{water_surface}
\end{center}
\end{figure}

Here, initially we examine the energetic stability of water molecules adsorbed 
on the  separated components, {\it viz.}: water on \slp\ (\water/\slp), and  on 
the amorphous a-\sio\ surface (\water/a-\sio). For the former system, we have 
considered two (energetically stable) \water/\slp\ configurations, 
Figs.\,\ref{water_surface}(a) and \ref{water_surface}(b); both recently proposed 
in the literature\,\cite{cai2015energetics,wang2016degradation}. We found that 
those two structures  are very close in energy, where we obtained \water\ 
adsorption energies of 0.18 and 0.21\,eV/molecule, respectively. Our total 
energy results  support the ``two leg'' geometry recently proposed by Wang {\it 
et al.}\,\cite{wang2016degradation}. At the equilibrium geometry, the molecular 
structure of  \water\  has been preserved, lying at 2.50~\AA\ from 
the \slp\ surface. 

There are several studies addressing the adsorption of water molecules on silica 
surfaces. The subject becomes more complicated and interesting for amorphous 
surfaces, since we have a very large number of energetically metastable  
configurations of \water\  on a-\sio. Here we have considered two  different 
geometries, {\it viz.}: (i) a \water\ molecule attached to the four-fold 
coordinated surface Si atom [\water/a-\sio\ in Fig.\ref{water_surface}(c)], and 
(ii)  \water\ dissociated geometry, giving rise to two  hydroxyls,  one composed 
by the oxygen and hydrogen atoms remnant from the adsorbed \water\ molecule, and 
another composed by a H atom, dissociated from the original \water\ molecule, 
bonded to the two-fold coordinate oxygen  atom of the a-\sio\ surface 
[OH/(OH)a-\sio\ in Fig.\,\ref{water_surface}(d)]. In (i) we find an adsorption 
energy of 1.19\,eV/molecule, with Si--O equilibrium bond length of 1.82\,\AA\ 
(close to the sum of the covalent radii of Si and O, 1.83\,\AA), indicating the 
formation of chemical bond between the \water\ molecule and the a-\sio\ surface. 
Our adsorption energy, and Si--O bond length results are in good agreement with 
the ones obtained by Zhao and Jing\,\cite{zhi2008structural}, for \water\ 
adsorbed on \sio\ clusters. On the other hand, we find that once the \water\ 
molecule is adsorbed on the a-\sio\ surface,  the formation of  hydroxyl groups 
[(ii)], \water/a-\sio$\rightarrow$OH/(OH)a-\sio, is an exothermic process by  
1.45\,eV. Indeed, the formation of OH groups on a-\sio\ has been supported by 
other theoretical studies 
\cite{walsh2000hydrolysis,mahadevan2008dissociative,lockwood2014proton}, and it 
is in  agreement with the experimental observation\,\cite{kim2003dissociation}. 
Those adsorption energy results  allow us to confirm the hydrophilic character 
of the a-\sio\ surface with respect to the \slp. In a recent experimental study, 
performed by Wood {\it et al.}\,\cite{wood2014effective}, the authors verified 
that the oxidation rate, of exfoliated black phosphorous, has been reduced by  
the presence of the (hydrophilic)  \sio\ surface. 

\begin{figure}[!ht]
\begin{center}
\includegraphics[width=8.5cm]{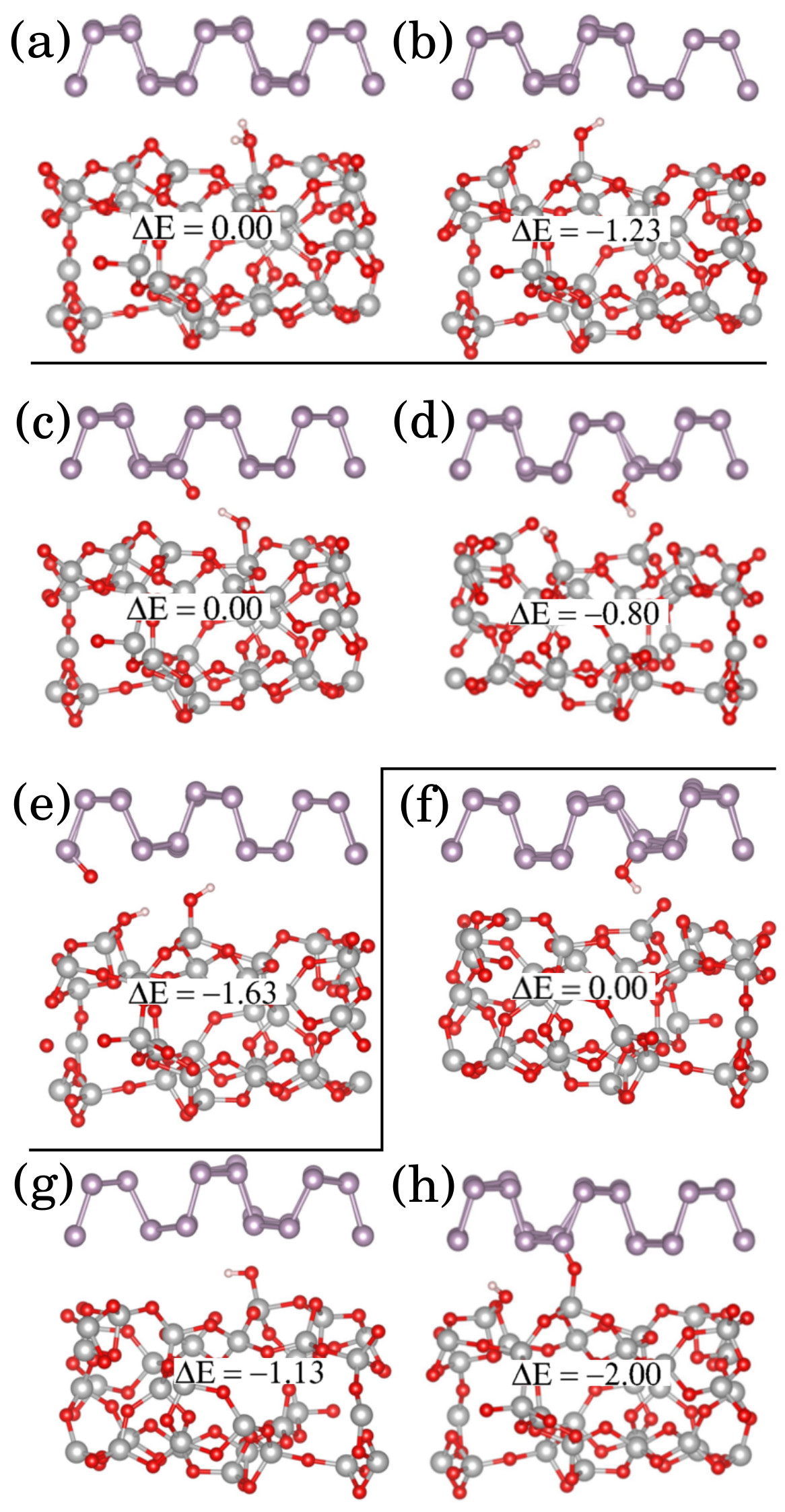}
\caption{Equilibrium geometries of \water\ molecules embedded in the 
\slp/a-\sio\ interface, undissociated (a), and dissociated formig OH groups (b).
Esquibrium geometries of \water\ molecules intercalated between the oxidized 
\slp\ and the a-\sio\ surface, undissociated (c), dissociate forming an OH 
group bonded to the \slp\ and another bonded to the a-\sio\ surface (d),  and 
forming  two OH groups bonded to the a-\sio\ surface (e).  Equilibrium geometry 
of an OH group bonded to the \slp\ (f), an OH group bonded to the a-\sio\ (g), 
and an oxygen atoms forming a P--O--Si bridge structure (h). The P, O, Si and H 
atoms are represented by light purple, red and gray color spheres.}
\label{OH}
\end{center}
\end{figure}

Next we examine the presence of \water\ at the \slp$-$a-\sio\ interface region. 
Within our calculation approach, the  interactions between the \water\ molecules 
are negligible; we are considering  lateral distances of about 13\,\AA\ between 
the molecules, which corresponds to a planar concentration of 
$5.6\times10^{13}$/cm$^2$. At such a low coverage regime, the adsorption energy 
of \slp\ on the \water/a-\sio\ surface [\slp/\water/a-\sio\ in 
Fig.\,\ref{OH}(a)] is practically the same as that obtained without \water\ 
molecules. We have considered two \slp/\water/a-\sio\ configurations, where we 
found $E^{\rm a}$ of 13.2 and 15.8\,meV/\AA$^2$, and the vertical distance 
between the \slp\ and the a-\sio\ surface increases to 3.47\,\AA\ (averaged 
value). 

Similarly to the \water/a-\sio\ systems discussed above, the formation of 
hydroxyl groups even at the presence of the \slp\ layer, 
\slp/\water/a-\sio$\rightarrow$\slp/OH/(OH)a-\sio\ [Fig.\,\ref{OH}(b)], is an 
exothermic process;  the total energy reduces by 1.23\,eV. The presence of 
hydroxyl groups at the \slp$-$a-\sio\ interface increases the  corrugation of 
the \slp\ by 0.14\,\AA, in comparison with the \slp/a-\sio\ system, 
0.08$\rightarrow$0.22\,\AA, where the \slp\ lies at 2.50\,\AA\ from the 
(OH)/(OH)a-\sio\ substrate (averaged values). Indeed, similar equilibrium 
geometry picture has been verified for graphene adsorbed on 
a-HfO$_2$, intercalated by \water\ molecules\,\cite{olson2015capacitive}.

In order to provide  a more complete picture of water in \slp/a-\sio,  we have 
also considered the presence of oxygen, (i) as a interstitial defect attached to 
the \slp, and (ii) forming  oxygenated \water, embedded between the phosphorene 
sheet and the a-\sio\ substrate\,\cite{wood2014effective}. For the interstitial 
oxygen, our binding energy, and equilibrium geometry results are in agreement 
with the ones of the most stable configuration predicted by Ziletti {\it et 
al.}\,\cite{ziletti2015oxygen}.

\begin{figure}[!ht]
\begin{center}
\includegraphics[width=8.5cm]{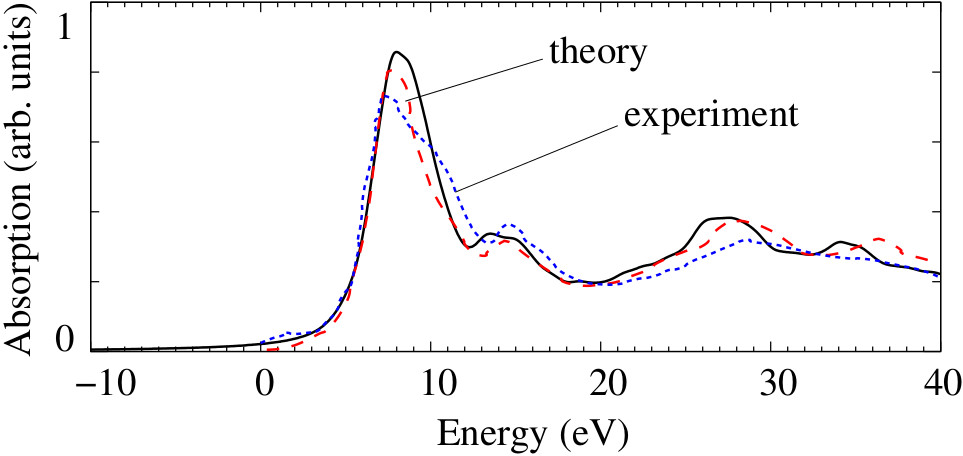}
\caption{Oxygen K-edge XANES spectrum obtained for the $\alpha$-quartz
geometry comparing with experimental and theoretical 
data\cite{taillefumier2002}.}
\label{quartz}
\end{center}
\end{figure}

The structural model, and the total energy differences in (i) are depicted in 
Figs.\,\ref{OH}(c)--\ref{OH}(e); by comparing the total energies, we found the 
latter (former) configuration as the most (less) stable one. Thus, indicating  
that, even at the presence of interstitial oxygen attached to  the \slp,  the 
remnant H atom from the \water\ molecule promotes the formation hydroxyls on the 
a-\sio\ surface, Fig.\,\ref{OH}(e), as we verified on the a-\sio\ pristine 
surface. In this case, there is an atomic rearrangement on the a-\sio\ surface, 
in order to keep the two-fold (four-fold) coordination of the surface O (Si)  
atoms. However, it is worth noting that due to the amorphous character of the 
surface, the formation of hydroxyl groups on  a-\sio\ will depend on the local 
geometry, and thus, although energetically less stable, we may find the 
formation of hydroxyl groups on the \slp, as shown in  Fig.\,\ref{OH}(d).

In (ii),  we find that dissociation of \watero, giving rise to hydroxyl groups 
attached to the \slp\ or a-\sio\ surface, is an exothermic process. Here we have 
considered a number of plausible configurations of OH at the \slp$-$a-\sio\  
interface. For an OH  group attached to the \slp\ sheet, Fig.\,\ref{OH}(f), we 
find a total energy release of 0.99\,eV/OH, when compared with the separated 
components, namely, pristine \slp/a-\sio\ and a \watero\ molecule. Further total 
energy comparisons indicate that the \slp/(OH)a-\sio\ configuration 
[Fig.\,\ref{OH}(g)] is more stable by 1.13\,eV. Whereas, again due to the 
amorphous character of the surface, we may find an energetically quite stable 
configuration composed by an oxygen atom (two fold coordinated) forming a 
P--O--Si bridge structure between the \slp\ and the a-\sio\ surface; while the 
remnant hydrogen atom forms an OH group with the a-\sio\ surface, 
Fig.\,\ref{OH}(h).

\begin{figure}[!ht]
\begin{center}
\includegraphics[width=7cm]{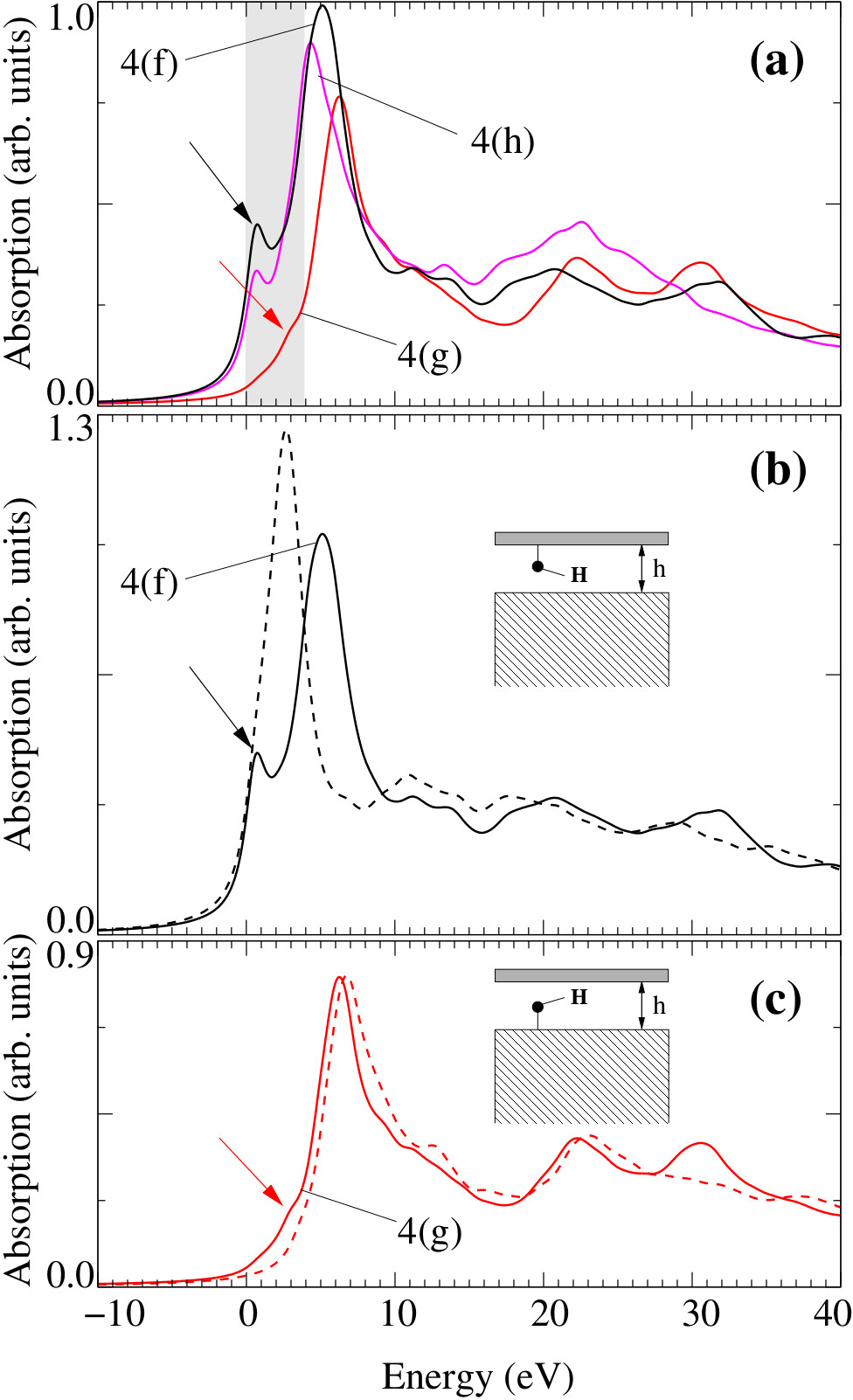}
\caption{Simulated oxygen K-edge XAS spectrum of the OH group for structural 
models depicted in Figs.\,\ref{OH}(f), \ref{OH}(g), and \ref{OH}(h) (a). 
O K-edge spectrum of the OH group of the structural models 
shown in Fig.\,\ref{OH}(f) and \ref{OH}(g). (b) at the equilibrium geometry 
(h=2.00\,\AA, solid-line) and increasing the vertical distance (h=4.00\,\AA, 
dashed-line), and (c) at the equilibrium geometry (h=2.82\,\AA, solid-line) and 
increasing the vertical distance (h=6.82\,\AA, dashed-line).}
\label{x-new}
\end{center}
\end{figure}

X-ray absorption spectroscopy (XAS) has been considered a quite suitable 
technique to provide informations about atomic and electronic structure near 
the probed element. Here, we  characterize the presence of OH groups  by 
performing a set of XAS simulations of the oxygen K-edge spectrum in  
\slp/a-\sio.

Initially, in order to verify the adequacy of our calculation procedure, we  
calculated the oxygen K-edge spectrum of   $\alpha$-quartz (SiO$_{2}$). Where, 
as shown in Fig.\,\ref{quartz} (solid lines), we find a good agreement with the 
previous theoretical/experimental findings\,\cite{taillefumier2002}. For the 
\slp/a-\sio, we examine   the K-edge spectrum of the oxygen atom of the OH 
group, as shown in Figs.\,\ref{OH}(f) and \ref{OH}(g), as well as the one 
forming the P-O-Si bridge structure presented in Fig.\,\ref{OH}(h).
Based on the PDOS 
results, shown in Fig.\,\ref{energies_level}(a), the final state in the 
O-1$s\rightarrow p$ transition should present a dominant contribution from the 
P-3$p$  orbitals of the  SLP; even upon the presence of a core hole in the  
(probed) oxygen atom. Thus, suggesting  a weak dependence between the O K-edge 
energy position and the configuration  of the oxygen atom in \slp/a-\sio. 
Indeed, for the three structural models, the XAS spectra present somewhat the 
same feature, with the O-1$s$ peak lying within the shaded region in  
Fig.\ref{x-new}(a). Indicated by arrows in Fig.\,\ref{x-new}(a), we verify that 
the O-1$s$ peak  is   more intense  for OH group bonded to the SLP [structural 
model shown in Fig.\,\ref{OH}(f)], and less intense for OH bonded to the \sio\ 
surface [Fig.\,\ref{OH}(g)]. Meanwhile, for O atom forming the P-O-Si bridge 
structure, Fig.\,\ref{OH}(h), we have an intermediate intensity.

Computational simulations can be a quite useful tool to help the interpretation 
of XAS spectra. For instance, we can simulate a set of configurations, which are 
experimentally non-accessible,  aiming to provide additional informations of  a 
given spectrum. In  Fig.\,\ref{x-new}(b), the solid line indicate  the original 
O K-edge spectrum of OH bonded to the SLP [Fig.\,\ref{OH}(f)]; while the dashed 
line corresponds to the O K-edge spectrum of the same OH configuration, but 
increasing the vertical distance (h) from 2.00\,\AA\ (equilibrium geometry) to 
4.00\,\AA. In this case, the role played by the a-\sio\ surface, to the XAS 
spectrum, has been reduced. We find that the two (original) O-1$s$ peaks reduce 
to a single one with higher intensity. Thus, indicating that (i) the a-\sio\ 
surface orbitals contribute to the final state of O K-edge spectrum of OH bonded 
to the SLP, and (ii) the interaction with the a-SiO$_{2}$\ surface  reduces the  
O-1$s$ binding energy. On the other hand, for OH group bonded to the a-\sio\ 
surface [Fig.\,\ref{OH}(g)], (i) the feature indicated by an arrow in 
Fig.\,\ref{x-new}(c) has been suppressed by increasing the vertical distance, h 
= 2.82 (equilibrium position) $\rightarrow$ 6.82\,\AA, and (ii) there is an 
increase on the O-1$s$ binding energy. Those results  can be attributed to the 
reduction of electronic contributions from the phosphorene to the final state of 
O K-edge spectrum.




\section{Summary}

In summary, based on first-principles calculations we investigate (i) the 
electronic and structural properties of a single layer phosphorene (\slp) 
adsorbed on the amorphous silicon oxide surface (\slp/a-\sio), and (ii) the 
incorporation of \water\ molecules in the \slp$-$a-\sio\ interface. In (i), we 
find that single layer phosphorene bonds to the amorphous \sio\ surface 
(\slp/a-\sio) mediated by vdW interactions. There are no chemical bonds between 
the SLP and the surface, even at the presence of oxygen vacancies ($\rm V_O$), 
which is a common intrinsic defect in a-\sio. The (amorphous) surface 
corrugation promotes a non-uniform charge density distribution, giving rise to 
electron-hole puddles in  \slp/a-\sio. Such  charge density puddles are 
strengthened upon the presence of $\rm V_O$, which may reduce the carrier 
mobility in the phosphorene layer. The $\rm V_O$ defect gives rise to an 
occupied level in the band gap of a-\sio; upon the formation of \slp/a-\sio, the 
defect level lies near the valence band maximum of the \slp. We estimate the 
band alignment of \slp/a-\sio, where we find a type-I band offset.  In (ii), we 
find an energetic preference for the formation of hydroxyl (OH) groups bonded to 
the pristine a-\sio\ surface, as well as upon the presence of the \slp, 
\slp/a-\sio.  However, due to the  amorphous structure of the surface, depending 
on the local geometry, we may find other energetically stable configurations. 
Indeed, we found  one  composed by an hydrogen atom bonded to the a-\sio\ 
surface and the  oxygen atom forming a P--O--Si bridge structure. Further x-ray 
absorption spectroscopy (XAS) simulations, of oxygen atom forming OH or P--O--Si 
bridge structure, reveal that the K-edge spectra present different features 
for each configuration.

\section*{Acknowledgments}

The authors acknowledge the support from Brazilian
 agencies CNPq, FAPES, FAPEMIG and CAPES; we thank also to CENAPAD-SP for 
computer time.
\section*{References}

\bibliography{references-iop.bib}

\begin{thebibliography}{55}%
\makeatletter
\providecommand \@ifxundefined [1]{%
 \@ifx{#1\undefined}
}%
\providecommand \@ifnum [1]{%
 \ifnum #1\expandafter \@firstoftwo
 \else \expandafter \@secondoftwo
 \fi
}%
\providecommand \@ifx [1]{%
 \ifx #1\expandafter \@firstoftwo
 \else \expandafter \@secondoftwo
 \fi
}%
\providecommand \natexlab [1]{#1}%
\providecommand \enquote  [1]{``#1''}%
\providecommand \bibnamefont  [1]{#1}%
\providecommand \bibfnamefont [1]{#1}%
\providecommand \citenamefont [1]{#1}%
\providecommand \href@noop [0]{\@secondoftwo}%
\providecommand \href [0]{\begingroup \@sanitize@url \@href}%
\providecommand \@href[1]{\@@startlink{#1}\@@href}%
\providecommand \@@href[1]{\endgroup#1\@@endlink}%
\providecommand \@sanitize@url [0]{\catcode `\\12\catcode `\$12\catcode
  `\&12\catcode `\#12\catcode `\^12\catcode `\_12\catcode `\%12\relax}%
\providecommand \@@startlink[1]{}%
\providecommand \@@endlink[0]{}%
\providecommand \url  [0]{\begingroup\@sanitize@url \@url }%
\providecommand \@url [1]{\endgroup\@href {#1}{\urlprefix }}%
\providecommand \urlprefix  [0]{URL }%
\providecommand \Eprint [0]{\href }%
\providecommand \doibase [0]{http://dx.doi.org/}%
\providecommand \selectlanguage [0]{\@gobble}%
\providecommand \bibinfo  [0]{\@secondoftwo}%
\providecommand \bibfield  [0]{\@secondoftwo}%
\providecommand \translation [1]{[#1]}%
\providecommand \BibitemOpen [0]{}%
\providecommand \bibitemStop [0]{}%
\providecommand \bibitemNoStop [0]{.\EOS\space}%
\providecommand \EOS [0]{\spacefactor3000\relax}%
\providecommand \BibitemShut  [1]{\csname bibitem#1\endcsname}%
\let\auto@bib@innerbib\@empty
\bibitem [{\citenamefont {Zhang}\ \emph {et~al.}(2014)\citenamefont {Zhang},
  \citenamefont {Yang}, \citenamefont {Xu}, \citenamefont {Wang}, \citenamefont
  {Li}, \citenamefont {Ghufran}, \citenamefont {Zhang}, \citenamefont {Yu},
  \citenamefont {Zhang}, \citenamefont {Qin} \emph
  {et~al.}}]{zhang2014extraordinary}%
  \BibitemOpen
  \bibfield  {author} {\bibinfo {author} {\bibfnamefont {S.}~\bibnamefont
  {Zhang}}, \bibinfo {author} {\bibfnamefont {J.}~\bibnamefont {Yang}},
  \bibinfo {author} {\bibfnamefont {R.}~\bibnamefont {Xu}}, \bibinfo {author}
  {\bibfnamefont {F.}~\bibnamefont {Wang}}, \bibinfo {author} {\bibfnamefont
  {W.}~\bibnamefont {Li}}, \bibinfo {author} {\bibfnamefont {M.}~\bibnamefont
  {Ghufran}}, \bibinfo {author} {\bibfnamefont {Y.-W.}\ \bibnamefont {Zhang}},
  \bibinfo {author} {\bibfnamefont {Z.}~\bibnamefont {Yu}}, \bibinfo {author}
  {\bibfnamefont {G.}~\bibnamefont {Zhang}}, \bibinfo {author} {\bibfnamefont
  {Q.}~\bibnamefont {Qin}},  \emph {et~al.},\ }\href@noop {} {\bibfield
  {journal} {\bibinfo  {journal} {ACS nano}\ }\textbf {\bibinfo {volume} {8}},\
  \bibinfo {pages} {9590} (\bibinfo {year} {2014})}\BibitemShut {NoStop}%
\bibitem [{\citenamefont {Tran}\ \emph {et~al.}(2014)\citenamefont {Tran},
  \citenamefont {Soklaski}, \citenamefont {Liang},\ and\ \citenamefont
  {Yang}}]{tran2014layer}%
  \BibitemOpen
  \bibfield  {author} {\bibinfo {author} {\bibfnamefont {V.}~\bibnamefont
  {Tran}}, \bibinfo {author} {\bibfnamefont {R.}~\bibnamefont {Soklaski}},
  \bibinfo {author} {\bibfnamefont {Y.}~\bibnamefont {Liang}}, \ and\ \bibinfo
  {author} {\bibfnamefont {L.}~\bibnamefont {Yang}},\ }\href@noop {} {\bibfield
   {journal} {\bibinfo  {journal} {Physical Review B}\ }\textbf {\bibinfo
  {volume} {89}},\ \bibinfo {pages} {235319} (\bibinfo {year}
  {2014})}\BibitemShut {NoStop}%
\bibitem [{\citenamefont {Qiao}\ \emph {et~al.}(2014)\citenamefont {Qiao},
  \citenamefont {Kong}, \citenamefont {Hu}, \citenamefont {Yang},\ and\
  \citenamefont {Ji}}]{qiao2014high}%
  \BibitemOpen
  \bibfield  {author} {\bibinfo {author} {\bibfnamefont {J.}~\bibnamefont
  {Qiao}}, \bibinfo {author} {\bibfnamefont {X.}~\bibnamefont {Kong}}, \bibinfo
  {author} {\bibfnamefont {Z.-X.}\ \bibnamefont {Hu}}, \bibinfo {author}
  {\bibfnamefont {F.}~\bibnamefont {Yang}}, \ and\ \bibinfo {author}
  {\bibfnamefont {W.}~\bibnamefont {Ji}},\ }\href@noop {} {\bibfield  {journal}
  {\bibinfo  {journal} {Nature communications}\ }\textbf {\bibinfo {volume}
  {5}} (\bibinfo {year} {2014})}\BibitemShut {NoStop}%
\bibitem [{\citenamefont {Liu}\ \emph {et~al.}(2014)\citenamefont {Liu},
  \citenamefont {Neal}, \citenamefont {Zhu}, \citenamefont {Luo}, \citenamefont
  {Xu}, \citenamefont {Tom{\'a}nek},\ and\ \citenamefont
  {Ye}}]{liu2014phosphorene}%
  \BibitemOpen
  \bibfield  {author} {\bibinfo {author} {\bibfnamefont {H.}~\bibnamefont
  {Liu}}, \bibinfo {author} {\bibfnamefont {A.~T.}\ \bibnamefont {Neal}},
  \bibinfo {author} {\bibfnamefont {Z.}~\bibnamefont {Zhu}}, \bibinfo {author}
  {\bibfnamefont {Z.}~\bibnamefont {Luo}}, \bibinfo {author} {\bibfnamefont
  {X.}~\bibnamefont {Xu}}, \bibinfo {author} {\bibfnamefont {D.}~\bibnamefont
  {Tom{\'a}nek}}, \ and\ \bibinfo {author} {\bibfnamefont {P.~D.}\ \bibnamefont
  {Ye}},\ }\href@noop {} {\bibfield  {journal} {\bibinfo  {journal} {ACS nano}\
  }\textbf {\bibinfo {volume} {8}},\ \bibinfo {pages} {4033} (\bibinfo {year}
  {2014})}\BibitemShut {NoStop}%
\bibitem [{\citenamefont {Li}\ \emph {et~al.}(2014)\citenamefont {Li},
  \citenamefont {Yu}, \citenamefont {Ye}, \citenamefont {Ge}, \citenamefont
  {Ou}, \citenamefont {Wu}, \citenamefont {Feng}, \citenamefont {Chen},\ and\
  \citenamefont {Zhang}}]{li2014black}%
  \BibitemOpen
  \bibfield  {author} {\bibinfo {author} {\bibfnamefont {L.}~\bibnamefont
  {Li}}, \bibinfo {author} {\bibfnamefont {Y.}~\bibnamefont {Yu}}, \bibinfo
  {author} {\bibfnamefont {G.~J.}\ \bibnamefont {Ye}}, \bibinfo {author}
  {\bibfnamefont {Q.}~\bibnamefont {Ge}}, \bibinfo {author} {\bibfnamefont
  {X.}~\bibnamefont {Ou}}, \bibinfo {author} {\bibfnamefont {H.}~\bibnamefont
  {Wu}}, \bibinfo {author} {\bibfnamefont {D.}~\bibnamefont {Feng}}, \bibinfo
  {author} {\bibfnamefont {X.~H.}\ \bibnamefont {Chen}}, \ and\ \bibinfo
  {author} {\bibfnamefont {Y.}~\bibnamefont {Zhang}},\ }\href@noop {}
  {\bibfield  {journal} {\bibinfo  {journal} {Nature nanotechnology}\ }\textbf
  {\bibinfo {volume} {9}},\ \bibinfo {pages} {372} (\bibinfo {year}
  {2014})}\BibitemShut {NoStop}%
\bibitem [{\citenamefont {Das}\ \emph {et~al.}(2014)\citenamefont {Das},
  \citenamefont {Demarteau},\ and\ \citenamefont {Roelofs}}]{das2014ambipolar}%
  \BibitemOpen
  \bibfield  {author} {\bibinfo {author} {\bibfnamefont {S.}~\bibnamefont
  {Das}}, \bibinfo {author} {\bibfnamefont {M.}~\bibnamefont {Demarteau}}, \
  and\ \bibinfo {author} {\bibfnamefont {A.}~\bibnamefont {Roelofs}},\
  }\href@noop {} {\bibfield  {journal} {\bibinfo  {journal} {ACS nano}\
  }\textbf {\bibinfo {volume} {8}},\ \bibinfo {pages} {11730} (\bibinfo {year}
  {2014})}\BibitemShut {NoStop}%
\bibitem [{\citenamefont {Zhu}\ \emph {et~al.}(2015)\citenamefont {Zhu},
  \citenamefont {Yogeesh}, \citenamefont {Yang}, \citenamefont {Aldave},
  \citenamefont {Kim}, \citenamefont {Sonde}, \citenamefont {Tao},
  \citenamefont {Lu},\ and\ \citenamefont {Akinwande}}]{zhu2015flexible}%
  \BibitemOpen
  \bibfield  {author} {\bibinfo {author} {\bibfnamefont {W.}~\bibnamefont
  {Zhu}}, \bibinfo {author} {\bibfnamefont {M.~N.}\ \bibnamefont {Yogeesh}},
  \bibinfo {author} {\bibfnamefont {S.}~\bibnamefont {Yang}}, \bibinfo {author}
  {\bibfnamefont {S.~H.}\ \bibnamefont {Aldave}}, \bibinfo {author}
  {\bibfnamefont {J.-S.}\ \bibnamefont {Kim}}, \bibinfo {author} {\bibfnamefont
  {S.}~\bibnamefont {Sonde}}, \bibinfo {author} {\bibfnamefont
  {L.}~\bibnamefont {Tao}}, \bibinfo {author} {\bibfnamefont {N.}~\bibnamefont
  {Lu}}, \ and\ \bibinfo {author} {\bibfnamefont {D.}~\bibnamefont
  {Akinwande}},\ }\href@noop {} {\bibfield  {journal} {\bibinfo  {journal}
  {Nano letters}\ }\textbf {\bibinfo {volume} {15}},\ \bibinfo {pages} {1883}
  (\bibinfo {year} {2015})}\BibitemShut {NoStop}%
\bibitem [{\citenamefont {Du}\ \emph {et~al.}(2014)\citenamefont {Du},
  \citenamefont {Liu}, \citenamefont {Deng},\ and\ \citenamefont
  {Ye}}]{du2014device}%
  \BibitemOpen
  \bibfield  {author} {\bibinfo {author} {\bibfnamefont {Y.}~\bibnamefont
  {Du}}, \bibinfo {author} {\bibfnamefont {H.}~\bibnamefont {Liu}}, \bibinfo
  {author} {\bibfnamefont {Y.}~\bibnamefont {Deng}}, \ and\ \bibinfo {author}
  {\bibfnamefont {P.~D.}\ \bibnamefont {Ye}},\ }\href@noop {} {\bibfield
  {journal} {\bibinfo  {journal} {ACS nano}\ }\textbf {\bibinfo {volume} {8}},\
  \bibinfo {pages} {10035} (\bibinfo {year} {2014})}\BibitemShut {NoStop}%
\bibitem [{\citenamefont {Du}\ \emph {et~al.}(2015)\citenamefont {Du},
  \citenamefont {Lin}, \citenamefont {Xu},\ and\ \citenamefont
  {Chu}}]{du2015recent}%
  \BibitemOpen
  \bibfield  {author} {\bibinfo {author} {\bibfnamefont {H.}~\bibnamefont
  {Du}}, \bibinfo {author} {\bibfnamefont {X.}~\bibnamefont {Lin}}, \bibinfo
  {author} {\bibfnamefont {Z.}~\bibnamefont {Xu}}, \ and\ \bibinfo {author}
  {\bibfnamefont {D.}~\bibnamefont {Chu}},\ }\href@noop {} {\bibfield
  {journal} {\bibinfo  {journal} {Journal of Materials Chemistry C}\ }\textbf
  {\bibinfo {volume} {3}},\ \bibinfo {pages} {8760} (\bibinfo {year}
  {2015})}\BibitemShut {NoStop}%
\bibitem [{\citenamefont {Verzhbitskiy}\ \emph {et~al.}(2015)\citenamefont
  {Verzhbitskiy}, \citenamefont {Carvalho}, \citenamefont {Rodin},
  \citenamefont {Koenig}, \citenamefont {Eda}, \citenamefont {Chen},
  \citenamefont {Neto},\ and\ \citenamefont
  {Ozyilmaz}}]{verzhbitskiy2015colossal}%
  \BibitemOpen
  \bibfield  {author} {\bibinfo {author} {\bibfnamefont {I.}~\bibnamefont
  {Verzhbitskiy}}, \bibinfo {author} {\bibfnamefont {A.}~\bibnamefont
  {Carvalho}}, \bibinfo {author} {\bibfnamefont {A.~S.}\ \bibnamefont {Rodin}},
  \bibinfo {author} {\bibfnamefont {S.~P.}\ \bibnamefont {Koenig}}, \bibinfo
  {author} {\bibfnamefont {G.}~\bibnamefont {Eda}}, \bibinfo {author}
  {\bibfnamefont {W.}~\bibnamefont {Chen}}, \bibinfo {author} {\bibfnamefont
  {A.~C.}\ \bibnamefont {Neto}}, \ and\ \bibinfo {author} {\bibfnamefont
  {B.}~\bibnamefont {Ozyilmaz}},\ }\href@noop {} {\bibfield  {journal}
  {\bibinfo  {journal} {ACS nano}\ }\textbf {\bibinfo {volume} {9}},\ \bibinfo
  {pages} {8070} (\bibinfo {year} {2015})}\BibitemShut {NoStop}%
\bibitem [{\citenamefont {Favron}\ \emph {et~al.}(2015)\citenamefont {Favron},
  \citenamefont {Gaufr{\`e}s}, \citenamefont {Fossard}, \citenamefont
  {Phaneuf-L’Heureux}, \citenamefont {Tang}, \citenamefont {L{\'e}vesque},
  \citenamefont {Loiseau}, \citenamefont {Leonelli}, \citenamefont
  {Francoeur},\ and\ \citenamefont {Martel}}]{favron2015photooxidation}%
  \BibitemOpen
  \bibfield  {author} {\bibinfo {author} {\bibfnamefont {A.}~\bibnamefont
  {Favron}}, \bibinfo {author} {\bibfnamefont {E.}~\bibnamefont {Gaufr{\`e}s}},
  \bibinfo {author} {\bibfnamefont {F.}~\bibnamefont {Fossard}}, \bibinfo
  {author} {\bibfnamefont {A.-L.}\ \bibnamefont {Phaneuf-L’Heureux}},
  \bibinfo {author} {\bibfnamefont {N.~Y.}\ \bibnamefont {Tang}}, \bibinfo
  {author} {\bibfnamefont {P.~L.}\ \bibnamefont {L{\'e}vesque}}, \bibinfo
  {author} {\bibfnamefont {A.}~\bibnamefont {Loiseau}}, \bibinfo {author}
  {\bibfnamefont {R.}~\bibnamefont {Leonelli}}, \bibinfo {author}
  {\bibfnamefont {S.}~\bibnamefont {Francoeur}}, \ and\ \bibinfo {author}
  {\bibfnamefont {R.}~\bibnamefont {Martel}},\ }\href@noop {} {\bibfield
  {journal} {\bibinfo  {journal} {Nature Materials}\ }\textbf {\bibinfo
  {volume} {14}},\ \bibinfo {pages} {826} (\bibinfo {year} {2015})}\BibitemShut
  {NoStop}%
\bibitem [{\citenamefont {Lu}\ \emph {et~al.}(2014)\citenamefont {Lu},
  \citenamefont {Nan}, \citenamefont {Hong}, \citenamefont {Chen},
  \citenamefont {Zhu}, \citenamefont {Liang}, \citenamefont {Ma}, \citenamefont
  {Ni}, \citenamefont {Jin},\ and\ \citenamefont {Zhang}}]{lu2014plasma}%
  \BibitemOpen
  \bibfield  {author} {\bibinfo {author} {\bibfnamefont {W.}~\bibnamefont
  {Lu}}, \bibinfo {author} {\bibfnamefont {H.}~\bibnamefont {Nan}}, \bibinfo
  {author} {\bibfnamefont {J.}~\bibnamefont {Hong}}, \bibinfo {author}
  {\bibfnamefont {Y.}~\bibnamefont {Chen}}, \bibinfo {author} {\bibfnamefont
  {C.}~\bibnamefont {Zhu}}, \bibinfo {author} {\bibfnamefont {Z.}~\bibnamefont
  {Liang}}, \bibinfo {author} {\bibfnamefont {X.}~\bibnamefont {Ma}}, \bibinfo
  {author} {\bibfnamefont {Z.}~\bibnamefont {Ni}}, \bibinfo {author}
  {\bibfnamefont {C.}~\bibnamefont {Jin}}, \ and\ \bibinfo {author}
  {\bibfnamefont {Z.}~\bibnamefont {Zhang}},\ }\href@noop {} {\bibfield
  {journal} {\bibinfo  {journal} {Nano Research}\ }\textbf {\bibinfo {volume}
  {7}},\ \bibinfo {pages} {853} (\bibinfo {year} {2014})}\BibitemShut {NoStop}%
\bibitem [{\citenamefont {Guo}\ \emph {et~al.}(2015)\citenamefont {Guo},
  \citenamefont {Zhang}, \citenamefont {Lu}, \citenamefont {Wang},
  \citenamefont {Tang}, \citenamefont {Shao}, \citenamefont {Sun},
  \citenamefont {Xie}, \citenamefont {Wang}, \citenamefont {Yu} \emph
  {et~al.}}]{guo2015black}%
  \BibitemOpen
  \bibfield  {author} {\bibinfo {author} {\bibfnamefont {Z.}~\bibnamefont
  {Guo}}, \bibinfo {author} {\bibfnamefont {H.}~\bibnamefont {Zhang}}, \bibinfo
  {author} {\bibfnamefont {S.}~\bibnamefont {Lu}}, \bibinfo {author}
  {\bibfnamefont {Z.}~\bibnamefont {Wang}}, \bibinfo {author} {\bibfnamefont
  {S.}~\bibnamefont {Tang}}, \bibinfo {author} {\bibfnamefont {J.}~\bibnamefont
  {Shao}}, \bibinfo {author} {\bibfnamefont {Z.}~\bibnamefont {Sun}}, \bibinfo
  {author} {\bibfnamefont {H.}~\bibnamefont {Xie}}, \bibinfo {author}
  {\bibfnamefont {H.}~\bibnamefont {Wang}}, \bibinfo {author} {\bibfnamefont
  {X.-F.}\ \bibnamefont {Yu}},  \emph {et~al.},\ }\href@noop {} {\bibfield
  {journal} {\bibinfo  {journal} {Advanced Functional Materials}\ }\textbf
  {\bibinfo {volume} {25}},\ \bibinfo {pages} {6996} (\bibinfo {year}
  {2015})}\BibitemShut {NoStop}%
\bibitem [{\citenamefont {Brent}\ \emph {et~al.}(2014)\citenamefont {Brent},
  \citenamefont {Savjani}, \citenamefont {Lewis}, \citenamefont {Haigh},
  \citenamefont {Lewis},\ and\ \citenamefont {O'Brien}}]{brent2014production}%
  \BibitemOpen
  \bibfield  {author} {\bibinfo {author} {\bibfnamefont {J.~R.}\ \bibnamefont
  {Brent}}, \bibinfo {author} {\bibfnamefont {N.}~\bibnamefont {Savjani}},
  \bibinfo {author} {\bibfnamefont {E.~A.}\ \bibnamefont {Lewis}}, \bibinfo
  {author} {\bibfnamefont {S.~J.}\ \bibnamefont {Haigh}}, \bibinfo {author}
  {\bibfnamefont {D.~J.}\ \bibnamefont {Lewis}}, \ and\ \bibinfo {author}
  {\bibfnamefont {P.}~\bibnamefont {O'Brien}},\ }\href@noop {} {\bibfield
  {journal} {\bibinfo  {journal} {Chemical Communications}\ }\textbf {\bibinfo
  {volume} {50}},\ \bibinfo {pages} {13338} (\bibinfo {year}
  {2014})}\BibitemShut {NoStop}%
\bibitem [{\citenamefont {Kang}\ \emph {et~al.}(2015)\citenamefont {Kang},
  \citenamefont {Wood}, \citenamefont {Wells}, \citenamefont {Lee},
  \citenamefont {Liu}, \citenamefont {Chen},\ and\ \citenamefont
  {Hersam}}]{kang2015solvent}%
  \BibitemOpen
  \bibfield  {author} {\bibinfo {author} {\bibfnamefont {J.}~\bibnamefont
  {Kang}}, \bibinfo {author} {\bibfnamefont {J.~D.}\ \bibnamefont {Wood}},
  \bibinfo {author} {\bibfnamefont {S.~A.}\ \bibnamefont {Wells}}, \bibinfo
  {author} {\bibfnamefont {J.-H.}\ \bibnamefont {Lee}}, \bibinfo {author}
  {\bibfnamefont {X.}~\bibnamefont {Liu}}, \bibinfo {author} {\bibfnamefont
  {K.-S.}\ \bibnamefont {Chen}}, \ and\ \bibinfo {author} {\bibfnamefont
  {M.~C.}\ \bibnamefont {Hersam}},\ }\href@noop {} {\bibfield  {journal}
  {\bibinfo  {journal} {ACS nano}\ }\textbf {\bibinfo {volume} {9}},\ \bibinfo
  {pages} {3596} (\bibinfo {year} {2015})}\BibitemShut {NoStop}%
\bibitem [{\citenamefont {Koenig}\ \emph {et~al.}(2014)\citenamefont {Koenig},
  \citenamefont {Doganov}, \citenamefont {Schmidt}, \citenamefont {Neto},\ and\
  \citenamefont {Oezyilmaz}}]{koenig2014electric}%
  \BibitemOpen
  \bibfield  {author} {\bibinfo {author} {\bibfnamefont {S.~P.}\ \bibnamefont
  {Koenig}}, \bibinfo {author} {\bibfnamefont {R.~A.}\ \bibnamefont {Doganov}},
  \bibinfo {author} {\bibfnamefont {H.}~\bibnamefont {Schmidt}}, \bibinfo
  {author} {\bibfnamefont {A.~C.}\ \bibnamefont {Neto}}, \ and\ \bibinfo
  {author} {\bibfnamefont {B.}~\bibnamefont {Oezyilmaz}},\ }\href@noop {}
  {\bibfield  {journal} {\bibinfo  {journal} {Applied Physics Letters}\
  }\textbf {\bibinfo {volume} {104}},\ \bibinfo {pages} {103106} (\bibinfo
  {year} {2014})}\BibitemShut {NoStop}%
\bibitem [{\citenamefont {Woomer}\ \emph {et~al.}(2015)\citenamefont {Woomer},
  \citenamefont {Farnsworth}, \citenamefont {Hu}, \citenamefont {Wells},
  \citenamefont {Donley},\ and\ \citenamefont
  {Warren}}]{woomer2015phosphorene}%
  \BibitemOpen
  \bibfield  {author} {\bibinfo {author} {\bibfnamefont {A.~H.}\ \bibnamefont
  {Woomer}}, \bibinfo {author} {\bibfnamefont {T.~W.}\ \bibnamefont
  {Farnsworth}}, \bibinfo {author} {\bibfnamefont {J.}~\bibnamefont {Hu}},
  \bibinfo {author} {\bibfnamefont {R.~A.}\ \bibnamefont {Wells}}, \bibinfo
  {author} {\bibfnamefont {C.~L.}\ \bibnamefont {Donley}}, \ and\ \bibinfo
  {author} {\bibfnamefont {S.~C.}\ \bibnamefont {Warren}},\ }\href@noop {}
  {\bibfield  {journal} {\bibinfo  {journal} {ACS nano}\ }\textbf {\bibinfo
  {volume} {9}},\ \bibinfo {pages} {8869} (\bibinfo {year} {2015})}\BibitemShut
  {NoStop}%
\bibitem [{\citenamefont {Island}\ \emph {et~al.}(2015)\citenamefont {Island},
  \citenamefont {Steele}, \citenamefont {van~der Zant},\ and\ \citenamefont
  {Castellanos-Gomez}}]{island2015environmental}%
  \BibitemOpen
  \bibfield  {author} {\bibinfo {author} {\bibfnamefont {J.~O.}\ \bibnamefont
  {Island}}, \bibinfo {author} {\bibfnamefont {G.~A.}\ \bibnamefont {Steele}},
  \bibinfo {author} {\bibfnamefont {H.~S.}\ \bibnamefont {van~der Zant}}, \
  and\ \bibinfo {author} {\bibfnamefont {A.}~\bibnamefont
  {Castellanos-Gomez}},\ }\href@noop {} {\bibfield  {journal} {\bibinfo
  {journal} {2D Materials}\ }\textbf {\bibinfo {volume} {2}},\ \bibinfo {pages}
  {011002} (\bibinfo {year} {2015})}\BibitemShut {NoStop}%
\bibitem [{\citenamefont {Ziletti}\ \emph {et~al.}(2015)\citenamefont
  {Ziletti}, \citenamefont {Carvalho}, \citenamefont {Campbell}, \citenamefont
  {Coker},\ and\ \citenamefont {Neto}}]{ziletti2015oxygen}%
  \BibitemOpen
  \bibfield  {author} {\bibinfo {author} {\bibfnamefont {A.}~\bibnamefont
  {Ziletti}}, \bibinfo {author} {\bibfnamefont {A.}~\bibnamefont {Carvalho}},
  \bibinfo {author} {\bibfnamefont {D.~K.}\ \bibnamefont {Campbell}}, \bibinfo
  {author} {\bibfnamefont {D.~F.}\ \bibnamefont {Coker}}, \ and\ \bibinfo
  {author} {\bibfnamefont {A.~C.}\ \bibnamefont {Neto}},\ }\href@noop {}
  {\bibfield  {journal} {\bibinfo  {journal} {Physical Review Letters}\
  }\textbf {\bibinfo {volume} {114}},\ \bibinfo {pages} {046801} (\bibinfo
  {year} {2015})}\BibitemShut {NoStop}%
\bibitem [{\citenamefont {Cai}\ \emph {et~al.}(2015)\citenamefont {Cai},
  \citenamefont {Ke}, \citenamefont {Zhang},\ and\ \citenamefont
  {Zhang}}]{cai2015energetics}%
  \BibitemOpen
  \bibfield  {author} {\bibinfo {author} {\bibfnamefont {Y.}~\bibnamefont
  {Cai}}, \bibinfo {author} {\bibfnamefont {Q.}~\bibnamefont {Ke}}, \bibinfo
  {author} {\bibfnamefont {G.}~\bibnamefont {Zhang}}, \ and\ \bibinfo {author}
  {\bibfnamefont {Y.-W.}\ \bibnamefont {Zhang}},\ }\href@noop {} {\bibfield
  {journal} {\bibinfo  {journal} {The Journal of Physical Chemistry C}\
  }\textbf {\bibinfo {volume} {119}},\ \bibinfo {pages} {3102} (\bibinfo {year}
  {2015})}\BibitemShut {NoStop}%
\bibitem [{\citenamefont {Wang}\ \emph {et~al.}(2016)\citenamefont {Wang},
  \citenamefont {Slough}, \citenamefont {Pandey},\ and\ \citenamefont
  {Karna}}]{wang2016degradation}%
  \BibitemOpen
  \bibfield  {author} {\bibinfo {author} {\bibfnamefont {G.}~\bibnamefont
  {Wang}}, \bibinfo {author} {\bibfnamefont {W.~J.}\ \bibnamefont {Slough}},
  \bibinfo {author} {\bibfnamefont {R.}~\bibnamefont {Pandey}}, \ and\ \bibinfo
  {author} {\bibfnamefont {S.~P.}\ \bibnamefont {Karna}},\ }\href@noop {}
  {\bibfield  {journal} {\bibinfo  {journal} {2D Materials}\ }\textbf {\bibinfo
  {volume} {3}},\ \bibinfo {pages} {025011} (\bibinfo {year}
  {2016})}\BibitemShut {NoStop}%
\bibitem [{\citenamefont {Wood}\ \emph {et~al.}(2014)\citenamefont {Wood},
  \citenamefont {Wells}, \citenamefont {Jariwala}, \citenamefont {Chen},
  \citenamefont {Cho}, \citenamefont {Sangwan}, \citenamefont {Liu},
  \citenamefont {Lauhon}, \citenamefont {Marks},\ and\ \citenamefont
  {Hersam}}]{wood2014effective}%
  \BibitemOpen
  \bibfield  {author} {\bibinfo {author} {\bibfnamefont {J.~D.}\ \bibnamefont
  {Wood}}, \bibinfo {author} {\bibfnamefont {S.~A.}\ \bibnamefont {Wells}},
  \bibinfo {author} {\bibfnamefont {D.}~\bibnamefont {Jariwala}}, \bibinfo
  {author} {\bibfnamefont {K.-S.}\ \bibnamefont {Chen}}, \bibinfo {author}
  {\bibfnamefont {E.}~\bibnamefont {Cho}}, \bibinfo {author} {\bibfnamefont
  {V.~K.}\ \bibnamefont {Sangwan}}, \bibinfo {author} {\bibfnamefont
  {X.}~\bibnamefont {Liu}}, \bibinfo {author} {\bibfnamefont {L.~J.}\
  \bibnamefont {Lauhon}}, \bibinfo {author} {\bibfnamefont {T.~J.}\
  \bibnamefont {Marks}}, \ and\ \bibinfo {author} {\bibfnamefont {M.~C.}\
  \bibnamefont {Hersam}},\ }\href@noop {} {\bibfield  {journal} {\bibinfo
  {journal} {Nano letters}\ }\textbf {\bibinfo {volume} {14}},\ \bibinfo
  {pages} {6964} (\bibinfo {year} {2014})}\BibitemShut {NoStop}%
\bibitem [{\citenamefont {Na}\ \emph {et~al.}(2014)\citenamefont {Na},
  \citenamefont {Lee}, \citenamefont {Lim}, \citenamefont {Hwang},
  \citenamefont {Kim}, \citenamefont {Choi},\ and\ \citenamefont
  {Song}}]{na2014few}%
  \BibitemOpen
  \bibfield  {author} {\bibinfo {author} {\bibfnamefont {J.}~\bibnamefont
  {Na}}, \bibinfo {author} {\bibfnamefont {Y.~T.}\ \bibnamefont {Lee}},
  \bibinfo {author} {\bibfnamefont {J.~A.}\ \bibnamefont {Lim}}, \bibinfo
  {author} {\bibfnamefont {D.~K.}\ \bibnamefont {Hwang}}, \bibinfo {author}
  {\bibfnamefont {G.-T.}\ \bibnamefont {Kim}}, \bibinfo {author} {\bibfnamefont
  {W.~K.}\ \bibnamefont {Choi}}, \ and\ \bibinfo {author} {\bibfnamefont
  {Y.-W.}\ \bibnamefont {Song}},\ }\href@noop {} {\bibfield  {journal}
  {\bibinfo  {journal} {ACS nano}\ }\textbf {\bibinfo {volume} {8}},\ \bibinfo
  {pages} {11753} (\bibinfo {year} {2014})}\BibitemShut {NoStop}%
\bibitem [{\citenamefont {Kim}\ \emph {et~al.}(2015)\citenamefont {Kim},
  \citenamefont {Liu}, \citenamefont {Zhu}, \citenamefont {Kim}, \citenamefont
  {Wu}, \citenamefont {Tao}, \citenamefont {Dodabalapur}, \citenamefont {Lai},\
  and\ \citenamefont {Akinwande}}]{kim2014toward}%
  \BibitemOpen
  \bibfield  {author} {\bibinfo {author} {\bibfnamefont {J.-S.}\ \bibnamefont
  {Kim}}, \bibinfo {author} {\bibfnamefont {Y.}~\bibnamefont {Liu}}, \bibinfo
  {author} {\bibfnamefont {W.}~\bibnamefont {Zhu}}, \bibinfo {author}
  {\bibfnamefont {S.}~\bibnamefont {Kim}}, \bibinfo {author} {\bibfnamefont
  {D.}~\bibnamefont {Wu}}, \bibinfo {author} {\bibfnamefont {L.}~\bibnamefont
  {Tao}}, \bibinfo {author} {\bibfnamefont {A.}~\bibnamefont {Dodabalapur}},
  \bibinfo {author} {\bibfnamefont {K.}~\bibnamefont {Lai}}, \ and\ \bibinfo
  {author} {\bibfnamefont {D.}~\bibnamefont {Akinwande}},\ }\href@noop {}
  {\bibfield  {journal} {\bibinfo  {journal} {Scientific Reports}\ }\textbf
  {\bibinfo {volume} {5}},\ \bibinfo {pages} {8989} (\bibinfo {year}
  {2015})}\BibitemShut {NoStop}%
\bibitem [{\citenamefont {Pei}\ \emph {et~al.}(2016)\citenamefont {Pei},
  \citenamefont {Gai}, \citenamefont {Yang}, \citenamefont {Wang},
  \citenamefont {Yu}, \citenamefont {Choi}, \citenamefont {Luther-Davies},\
  and\ \citenamefont {Lu}}]{pei2016producing}%
  \BibitemOpen
  \bibfield  {author} {\bibinfo {author} {\bibfnamefont {J.}~\bibnamefont
  {Pei}}, \bibinfo {author} {\bibfnamefont {X.}~\bibnamefont {Gai}}, \bibinfo
  {author} {\bibfnamefont {J.}~\bibnamefont {Yang}}, \bibinfo {author}
  {\bibfnamefont {X.}~\bibnamefont {Wang}}, \bibinfo {author} {\bibfnamefont
  {Z.}~\bibnamefont {Yu}}, \bibinfo {author} {\bibfnamefont {D.-Y.}\
  \bibnamefont {Choi}}, \bibinfo {author} {\bibfnamefont {B.}~\bibnamefont
  {Luther-Davies}}, \ and\ \bibinfo {author} {\bibfnamefont {Y.}~\bibnamefont
  {Lu}},\ }\href@noop {} {\bibfield  {journal} {\bibinfo  {journal} {Nature
  communications}\ }\textbf {\bibinfo {volume} {7}} (\bibinfo {year}
  {2016})}\BibitemShut {NoStop}%
\bibitem [{\citenamefont {Kresse}\ and\ \citenamefont
  {Hafner}(1993{\natexlab{a}})}]{kresse1}%
  \BibitemOpen
  \bibfield  {author} {\bibinfo {author} {\bibfnamefont {G.}~\bibnamefont
  {Kresse}}\ and\ \bibinfo {author} {\bibfnamefont {J.}~\bibnamefont
  {Hafner}},\ }\href@noop {} {\bibfield  {journal} {\bibinfo  {journal} {Phys.
  Rev. B}\ }\textbf {\bibinfo {volume} {47}},\ \bibinfo {pages} {558} (\bibinfo
  {year} {1993}{\natexlab{a}})}\BibitemShut {NoStop}%
\bibitem [{\citenamefont {Kresse}\ and\ \citenamefont
  {Hafner}(1993{\natexlab{b}})}]{kresse2}%
  \BibitemOpen
  \bibfield  {author} {\bibinfo {author} {\bibfnamefont {G.}~\bibnamefont
  {Kresse}}\ and\ \bibinfo {author} {\bibfnamefont {J.}~\bibnamefont
  {Hafner}},\ }\href@noop {} {\bibfield  {journal} {\bibinfo  {journal} {Phys.
  Rev. B}\ }\textbf {\bibinfo {volume} {48}},\ \bibinfo {pages} {13115}
  (\bibinfo {year} {1993}{\natexlab{b}})}\BibitemShut {NoStop}%
\bibitem [{\citenamefont {Kresse}\ and\ \citenamefont
  {Furthmuller}(1996)}]{kresse3}%
  \BibitemOpen
  \bibfield  {author} {\bibinfo {author} {\bibfnamefont {G.}~\bibnamefont
  {Kresse}}\ and\ \bibinfo {author} {\bibfnamefont {J.}~\bibnamefont
  {Furthmuller}},\ }\href@noop {} {\bibfield  {journal} {\bibinfo  {journal}
  {Comput. Mater. Sci.}\ }\textbf {\bibinfo {volume} {6}},\ \bibinfo {pages}
  {15} (\bibinfo {year} {1996})}\BibitemShut {NoStop}%
\bibitem [{\citenamefont {Scopel}\ \emph {et~al.}(2008)\citenamefont {Scopel},
  \citenamefont {{da Silva}},\ and\ \citenamefont {Fazzio}}]{scopelPRB2008}%
  \BibitemOpen
  \bibfield  {author} {\bibinfo {author} {\bibfnamefont {W.~L.}\ \bibnamefont
  {Scopel}}, \bibinfo {author} {\bibfnamefont {A.~J.~R.}\ \bibnamefont {{da
  Silva}}}, \ and\ \bibinfo {author} {\bibfnamefont {A.}~\bibnamefont
  {Fazzio}},\ }\href@noop {} {\bibfield  {journal} {\bibinfo  {journal} {Phys.
  Rev. B}\ }\textbf {\bibinfo {volume} {77}},\ \bibinfo {pages} {172101}
  (\bibinfo {year} {2008})}\BibitemShut {NoStop}%
\bibitem [{\citenamefont {Bl{\"o}chl}(1994)}]{bl1994p}%
  \BibitemOpen
  \bibfield  {author} {\bibinfo {author} {\bibfnamefont {P.~E.}\ \bibnamefont
  {Bl{\"o}chl}},\ }\href@noop {} {\bibfield  {journal} {\bibinfo  {journal}
  {Phys Rev B}\ }\textbf {\bibinfo {volume} {50}},\ \bibinfo {pages} {17953}
  (\bibinfo {year} {1994})}\BibitemShut {NoStop}%
\bibitem [{\citenamefont {Kresse}\ and\ \citenamefont
  {Joubert}(1999)}]{kresse1999ultrasoft}%
  \BibitemOpen
  \bibfield  {author} {\bibinfo {author} {\bibfnamefont {G.}~\bibnamefont
  {Kresse}}\ and\ \bibinfo {author} {\bibfnamefont {D.}~\bibnamefont
  {Joubert}},\ }\href@noop {} {\bibfield  {journal} {\bibinfo  {journal}
  {Physical Review B}\ }\textbf {\bibinfo {volume} {59}},\ \bibinfo {pages}
  {1758} (\bibinfo {year} {1999})}\BibitemShut {NoStop}%
\bibitem [{\citenamefont {Dion}\ \emph {et~al.}(2004)\citenamefont {Dion},
  \citenamefont {Rydberg}, \citenamefont {Schr{\"o}der}, \citenamefont
  {Langreth},\ and\ \citenamefont {Lundqvist}}]{dion2004van}%
  \BibitemOpen
  \bibfield  {author} {\bibinfo {author} {\bibfnamefont {M.}~\bibnamefont
  {Dion}}, \bibinfo {author} {\bibfnamefont {H.}~\bibnamefont {Rydberg}},
  \bibinfo {author} {\bibfnamefont {E.}~\bibnamefont {Schr{\"o}der}}, \bibinfo
  {author} {\bibfnamefont {D.~C.}\ \bibnamefont {Langreth}}, \ and\ \bibinfo
  {author} {\bibfnamefont {B.~I.}\ \bibnamefont {Lundqvist}},\ }\href@noop {}
  {\bibfield  {journal} {\bibinfo  {journal} {Physical review letters}\
  }\textbf {\bibinfo {volume} {92}},\ \bibinfo {pages} {246401} (\bibinfo
  {year} {2004})}\BibitemShut {NoStop}%
\bibitem [{\citenamefont {Klime{\v{s}}}\ \emph {et~al.}(2011)\citenamefont
  {Klime{\v{s}}}, \citenamefont {Bowler},\ and\ \citenamefont
  {Michaelides}}]{klimevs2011van}%
  \BibitemOpen
  \bibfield  {author} {\bibinfo {author} {\bibfnamefont {J.}~\bibnamefont
  {Klime{\v{s}}}}, \bibinfo {author} {\bibfnamefont {D.~R.}\ \bibnamefont
  {Bowler}}, \ and\ \bibinfo {author} {\bibfnamefont {A.}~\bibnamefont
  {Michaelides}},\ }\href@noop {} {\bibfield  {journal} {\bibinfo  {journal}
  {Physical Review B}\ }\textbf {\bibinfo {volume} {83}},\ \bibinfo {pages}
  {195131} (\bibinfo {year} {2011})}\BibitemShut {NoStop}%
\bibitem [{\citenamefont {J.~Heyd}\ and\ \citenamefont
  {Ernzerhof}(2006)}]{ernzerbhof2006}%
  \BibitemOpen
  \bibfield  {author} {\bibinfo {author} {\bibfnamefont {G.~E.~S.}\
  \bibnamefont {J.~Heyd}}\ and\ \bibinfo {author} {\bibfnamefont
  {M.}~\bibnamefont {Ernzerhof}},\ }\href@noop {} {\bibfield  {journal}
  {\bibinfo  {journal} {J. Chem. Phys.}\ }\textbf {\bibinfo {volume} {124}},\
  \bibinfo {pages} {219906} (\bibinfo {year} {2006})}\BibitemShut {NoStop}%
\bibitem [{\citenamefont {Brown}\ and\ \citenamefont
  {Rundqvist}(1965)}]{brown1965refinement}%
  \BibitemOpen
  \bibfield  {author} {\bibinfo {author} {\bibfnamefont {A.}~\bibnamefont
  {Brown}}\ and\ \bibinfo {author} {\bibfnamefont {S.}~\bibnamefont
  {Rundqvist}},\ }\href@noop {} {\bibfield  {journal} {\bibinfo  {journal}
  {Acta Crystallographica}\ }\textbf {\bibinfo {volume} {19}},\ \bibinfo
  {pages} {684} (\bibinfo {year} {1965})}\BibitemShut {NoStop}%
\bibitem [{\citenamefont {Miwa}\ \emph {et~al.}(2011)\citenamefont {Miwa},
  \citenamefont {Schmidt}, \citenamefont {Scopel},\ and\ \citenamefont
  {Fazzio}}]{miwaAPL2011}%
  \BibitemOpen
  \bibfield  {author} {\bibinfo {author} {\bibfnamefont {R.~H.}\ \bibnamefont
  {Miwa}}, \bibinfo {author} {\bibfnamefont {T.~M.}\ \bibnamefont {Schmidt}},
  \bibinfo {author} {\bibfnamefont {W.~L.}\ \bibnamefont {Scopel}}, \ and\
  \bibinfo {author} {\bibfnamefont {A.}~\bibnamefont {Fazzio}},\ }\href@noop {}
  {\bibfield  {journal} {\bibinfo  {journal} {Appl. Phys. Lett.}\ }\textbf
  {\bibinfo {volume} {99}},\ \bibinfo {pages} {163108} (\bibinfo {year}
  {2011})}\BibitemShut {NoStop}%
\bibitem [{\citenamefont {Scopel}\ \emph {et~al.}(2015)\citenamefont {Scopel},
  \citenamefont {Miwa}, \citenamefont {Schmidt},\ and\ \citenamefont
  {Venezuela}}]{scopel2015mos2}%
  \BibitemOpen
  \bibfield  {author} {\bibinfo {author} {\bibfnamefont {W.}~\bibnamefont
  {Scopel}}, \bibinfo {author} {\bibfnamefont {R.}~\bibnamefont {Miwa}},
  \bibinfo {author} {\bibfnamefont {T.}~\bibnamefont {Schmidt}}, \ and\
  \bibinfo {author} {\bibfnamefont {P.}~\bibnamefont {Venezuela}},\ }\href@noop
  {} {\bibfield  {journal} {\bibinfo  {journal} {Journal of Applied Physics}\
  }\textbf {\bibinfo {volume} {117}},\ \bibinfo {pages} {194303} (\bibinfo
  {year} {2015})}\BibitemShut {NoStop}%
\bibitem [{\citenamefont {Ishigami}\ \emph {et~al.}(2007)\citenamefont
  {Ishigami}, \citenamefont {Chen}, \citenamefont {Cullen}, \citenamefont
  {Fuhrer},\ and\ \citenamefont {Williams}}]{ishigamiNanoLett2007}%
  \BibitemOpen
  \bibfield  {author} {\bibinfo {author} {\bibfnamefont {M.}~\bibnamefont
  {Ishigami}}, \bibinfo {author} {\bibfnamefont {J.~H.}\ \bibnamefont {Chen}},
  \bibinfo {author} {\bibfnamefont {W.~G.}\ \bibnamefont {Cullen}}, \bibinfo
  {author} {\bibfnamefont {M.~S.}\ \bibnamefont {Fuhrer}}, \ and\ \bibinfo
  {author} {\bibfnamefont {E.~D.}\ \bibnamefont {Williams}},\ }\href@noop {}
  {\bibfield  {journal} {\bibinfo  {journal} {Nano Lett.}\ }\textbf {\bibinfo
  {volume} {7}},\ \bibinfo {pages} {1643} (\bibinfo {year} {2007})}\BibitemShut
  {NoStop}%
\bibitem [{\citenamefont {Sinitskii}\ \emph {et~al.}(2010)\citenamefont
  {Sinitskii}, \citenamefont {V.Kosynkin}, \citenamefont {Dimiev},\ and\
  \citenamefont {J.~M}}]{sinitskiiACSNano2010}%
  \BibitemOpen
  \bibfield  {author} {\bibinfo {author} {\bibfnamefont {A.}~\bibnamefont
  {Sinitskii}}, \bibinfo {author} {\bibfnamefont {D.}~\bibnamefont
  {V.Kosynkin}}, \bibinfo {author} {\bibfnamefont {A.}~\bibnamefont {Dimiev}},
  \ and\ \bibinfo {author} {\bibfnamefont {T.}~\bibnamefont {J.~M}},\
  }\href@noop {} {\bibfield  {journal} {\bibinfo  {journal} {ACS Nano}\
  }\textbf {\bibinfo {volume} {4}},\ \bibinfo {pages} {3095} (\bibinfo {year}
  {2010})}\BibitemShut {NoStop}%
\bibitem [{\citenamefont {Martin}\ \emph {et~al.}(2008)\citenamefont {Martin},
  \citenamefont {Akerman}, \citenamefont {Ulbricht}, \citenamefont {Lohmann},
  \citenamefont {Smet},\ and\ \citenamefont {{Von
  Klitzing}}}]{martinNatPhys2008}%
  \BibitemOpen
  \bibfield  {author} {\bibinfo {author} {\bibfnamefont {J.}~\bibnamefont
  {Martin}}, \bibinfo {author} {\bibfnamefont {N.}~\bibnamefont {Akerman}},
  \bibinfo {author} {\bibfnamefont {G.}~\bibnamefont {Ulbricht}}, \bibinfo
  {author} {\bibfnamefont {T.}~\bibnamefont {Lohmann}}, \bibinfo {author}
  {\bibfnamefont {J.~H.}\ \bibnamefont {Smet}}, \ and\ \bibinfo {author}
  {\bibfnamefont {K.}~\bibnamefont {{Von Klitzing}}},\ }\href@noop {}
  {\bibfield  {journal} {\bibinfo  {journal} {Nat. Phys.}\ }\textbf {\bibinfo
  {volume} {4}},\ \bibinfo {pages} {144} (\bibinfo {year} {2008})}\BibitemShut
  {NoStop}%
\bibitem [{\citenamefont {Bader}(1990)}]{bader1}%
  \BibitemOpen
  \bibfield  {author} {\bibinfo {author} {\bibfnamefont {R.~F.~W.}\
  \bibnamefont {Bader}},\ }\href@noop {} {\emph {\bibinfo {title} {{Atoms in
  Molecules: A Quantum Theory}}}}\ (\bibinfo  {publisher} {Oxford University
  Press},\ \bibinfo {address} {New York},\ \bibinfo {year} {1990})\BibitemShut
  {NoStop}%
\bibitem [{\citenamefont {Tang}\ \emph {et~al.}(2009)\citenamefont {Tang},
  \citenamefont {Sanville},\ and\ \citenamefont {Henkelman}}]{bader2}%
  \BibitemOpen
  \bibfield  {author} {\bibinfo {author} {\bibfnamefont {W.}~\bibnamefont
  {Tang}}, \bibinfo {author} {\bibfnamefont {E.}~\bibnamefont {Sanville}}, \
  and\ \bibinfo {author} {\bibfnamefont {G.}~\bibnamefont {Henkelman}},\
  }\href@noop {} {\bibfield  {journal} {\bibinfo  {journal} {J. Phys. :
  Condens. Matter}\ }\textbf {\bibinfo {volume} {21}},\ \bibinfo {pages}
  {084204} (\bibinfo {year} {2009})}\BibitemShut {NoStop}%
\bibitem [{\citenamefont {Souza}\ \emph {et~al.}(2016)\citenamefont {Souza},
  \citenamefont {Scopel},\ and\ \citenamefont {Miwa}}]{souza2016switchable}%
  \BibitemOpen
  \bibfield  {author} {\bibinfo {author} {\bibfnamefont {E.~S.}\ \bibnamefont
  {Souza}}, \bibinfo {author} {\bibfnamefont {W.~L.}\ \bibnamefont {Scopel}}, \
  and\ \bibinfo {author} {\bibfnamefont {R.}~\bibnamefont {Miwa}},\ }\href@noop
  {} {\bibfield  {journal} {\bibinfo  {journal} {Physical Review B}\ }\textbf
  {\bibinfo {volume} {93}},\ \bibinfo {pages} {235308} (\bibinfo {year}
  {2016})}\BibitemShut {NoStop}%
\bibitem [{vac()}]{vacuum-level}%
  \BibitemOpen
  \href@noop {} {}\bibinfo {note} {Here, the vacuum level was obtained by
  computing the planar-averaged electrostatic potential perpendicularly to the
  surface plane.}\BibitemShut {Stop}%
\bibitem [{\citenamefont {Cai}\ \emph {et~al.}(2014)\citenamefont {Cai},
  \citenamefont {Zhang},\ and\ \citenamefont {Zhang}}]{cai2014layer}%
  \BibitemOpen
  \bibfield  {author} {\bibinfo {author} {\bibfnamefont {Y.}~\bibnamefont
  {Cai}}, \bibinfo {author} {\bibfnamefont {G.}~\bibnamefont {Zhang}}, \ and\
  \bibinfo {author} {\bibfnamefont {Y.-W.}\ \bibnamefont {Zhang}},\ }\href@noop
  {} {\bibfield  {journal} {\bibinfo  {journal} {Scientific reports}\ }\textbf
  {\bibinfo {volume} {4}} (\bibinfo {year} {2014})}\BibitemShut {NoStop}%
\bibitem [{\citenamefont {Srivastava}\ \emph {et~al.}(2015)\citenamefont
  {Srivastava}, \citenamefont {Hembram}, \citenamefont {Mizuseki},
  \citenamefont {Lee}, \citenamefont {Han},\ and\ \citenamefont
  {Kim}}]{srivastava2015tuning}%
  \BibitemOpen
  \bibfield  {author} {\bibinfo {author} {\bibfnamefont {P.}~\bibnamefont
  {Srivastava}}, \bibinfo {author} {\bibfnamefont {K.~P.}\ \bibnamefont
  {Hembram}}, \bibinfo {author} {\bibfnamefont {H.}~\bibnamefont {Mizuseki}},
  \bibinfo {author} {\bibfnamefont {K.-R.}\ \bibnamefont {Lee}}, \bibinfo
  {author} {\bibfnamefont {S.~S.}\ \bibnamefont {Han}}, \ and\ \bibinfo
  {author} {\bibfnamefont {S.}~\bibnamefont {Kim}},\ }\href@noop {} {\bibfield
  {journal} {\bibinfo  {journal} {The Journal of Physical Chemistry C}\
  }\textbf {\bibinfo {volume} {119}},\ \bibinfo {pages} {6530} (\bibinfo {year}
  {2015})}\BibitemShut {NoStop}%
\bibitem [{dip()}]{dipole}%
  \BibitemOpen
  \href@noop {} {}\bibinfo {note} {Since there are no chemical bonding at the
  SLP--a-SiO$_2$ interface region, the dipole effect on the band alignment
  should be small}\BibitemShut {NoStop}%
\bibitem [{\citenamefont {Sushko}\ \emph {et~al.}(2005)\citenamefont {Sushko},
  \citenamefont {Mukhopadhyay}, \citenamefont {Stoneham},\ and\ \citenamefont
  {Shluger}}]{sushko2005oxygen}%
  \BibitemOpen
  \bibfield  {author} {\bibinfo {author} {\bibfnamefont {P.}~\bibnamefont
  {Sushko}}, \bibinfo {author} {\bibfnamefont {S.}~\bibnamefont
  {Mukhopadhyay}}, \bibinfo {author} {\bibfnamefont {A.}~\bibnamefont
  {Stoneham}}, \ and\ \bibinfo {author} {\bibfnamefont {A.}~\bibnamefont
  {Shluger}},\ }\href@noop {} {\bibfield  {journal} {\bibinfo  {journal}
  {Microelectronic engineering}\ }\textbf {\bibinfo {volume} {80}},\ \bibinfo
  {pages} {292} (\bibinfo {year} {2005})}\BibitemShut {NoStop}%
\bibitem [{\citenamefont {Zhi}\ \emph {et~al.}(2008)\citenamefont {Zhi},
  \citenamefont {Zhao}, \citenamefont {Guo},\ and\ \citenamefont
  {Jing}}]{zhi2008structural}%
  \BibitemOpen
  \bibfield  {author} {\bibinfo {author} {\bibfnamefont {L.-l.}\ \bibnamefont
  {Zhi}}, \bibinfo {author} {\bibfnamefont {G.-f.}\ \bibnamefont {Zhao}},
  \bibinfo {author} {\bibfnamefont {L.-j.}\ \bibnamefont {Guo}}, \ and\
  \bibinfo {author} {\bibfnamefont {Q.}~\bibnamefont {Jing}},\ }\href@noop {}
  {\bibfield  {journal} {\bibinfo  {journal} {Physical Review B}\ }\textbf
  {\bibinfo {volume} {77}},\ \bibinfo {pages} {235435} (\bibinfo {year}
  {2008})}\BibitemShut {NoStop}%
\bibitem [{\citenamefont {Walsh}\ \emph {et~al.}(2000)\citenamefont {Walsh},
  \citenamefont {Wilson},\ and\ \citenamefont {Sutton}}]{walsh2000hydrolysis}%
  \BibitemOpen
  \bibfield  {author} {\bibinfo {author} {\bibfnamefont {T.~R.}\ \bibnamefont
  {Walsh}}, \bibinfo {author} {\bibfnamefont {M.}~\bibnamefont {Wilson}}, \
  and\ \bibinfo {author} {\bibfnamefont {A.~P.}\ \bibnamefont {Sutton}},\
  }\href@noop {} {\bibfield  {journal} {\bibinfo  {journal} {Journal of
  Chemical Physics}\ }\textbf {\bibinfo {volume} {113}},\ \bibinfo {pages}
  {9191} (\bibinfo {year} {2000})}\BibitemShut {NoStop}%
\bibitem [{\citenamefont {Mahadevan}\ and\ \citenamefont
  {Garofalini}(2008)}]{mahadevan2008dissociative}%
  \BibitemOpen
  \bibfield  {author} {\bibinfo {author} {\bibfnamefont {T.}~\bibnamefont
  {Mahadevan}}\ and\ \bibinfo {author} {\bibfnamefont {S.}~\bibnamefont
  {Garofalini}},\ }\href@noop {} {\bibfield  {journal} {\bibinfo  {journal}
  {The Journal of Physical Chemistry C}\ }\textbf {\bibinfo {volume} {112}},\
  \bibinfo {pages} {1507} (\bibinfo {year} {2008})}\BibitemShut {NoStop}%
\bibitem [{\citenamefont {Lockwood}\ and\ \citenamefont
  {Garofalini}(2014)}]{lockwood2014proton}%
  \BibitemOpen
  \bibfield  {author} {\bibinfo {author} {\bibfnamefont {G.~K.}\ \bibnamefont
  {Lockwood}}\ and\ \bibinfo {author} {\bibfnamefont {S.~H.}\ \bibnamefont
  {Garofalini}},\ }\href@noop {} {\bibfield  {journal} {\bibinfo  {journal}
  {The Journal of Physical Chemistry C}\ }\textbf {\bibinfo {volume} {118}},\
  \bibinfo {pages} {29750} (\bibinfo {year} {2014})}\BibitemShut {NoStop}%
\bibitem [{\citenamefont {Kim}\ \emph {et~al.}(2003)\citenamefont {Kim},
  \citenamefont {Wei}, \citenamefont {Stultz},\ and\ \citenamefont
  {Goodman}}]{kim2003dissociation}%
  \BibitemOpen
  \bibfield  {author} {\bibinfo {author} {\bibfnamefont {Y.}~\bibnamefont
  {Kim}}, \bibinfo {author} {\bibfnamefont {T.}~\bibnamefont {Wei}}, \bibinfo
  {author} {\bibfnamefont {J.}~\bibnamefont {Stultz}}, \ and\ \bibinfo {author}
  {\bibfnamefont {D.}~\bibnamefont {Goodman}},\ }\href@noop {} {\bibfield
  {journal} {\bibinfo  {journal} {Langmuir}\ }\textbf {\bibinfo {volume}
  {19}},\ \bibinfo {pages} {1140} (\bibinfo {year} {2003})}\BibitemShut
  {NoStop}%
\bibitem [{\citenamefont {Olson}\ \emph {et~al.}(2015)\citenamefont {Olson},
  \citenamefont {Ma}, \citenamefont {Sun}, \citenamefont {Ebrish},
  \citenamefont {Haratipour}, \citenamefont {Min}, \citenamefont {Aluru},\ and\
  \citenamefont {Koester}}]{olson2015capacitive}%
  \BibitemOpen
  \bibfield  {author} {\bibinfo {author} {\bibfnamefont {E.~J.}\ \bibnamefont
  {Olson}}, \bibinfo {author} {\bibfnamefont {R.}~\bibnamefont {Ma}}, \bibinfo
  {author} {\bibfnamefont {T.}~\bibnamefont {Sun}}, \bibinfo {author}
  {\bibfnamefont {M.~A.}\ \bibnamefont {Ebrish}}, \bibinfo {author}
  {\bibfnamefont {N.}~\bibnamefont {Haratipour}}, \bibinfo {author}
  {\bibfnamefont {K.}~\bibnamefont {Min}}, \bibinfo {author} {\bibfnamefont
  {N.~R.}\ \bibnamefont {Aluru}}, \ and\ \bibinfo {author} {\bibfnamefont
  {S.~J.}\ \bibnamefont {Koester}},\ }\href@noop {} {\bibfield  {journal}
  {\bibinfo  {journal} {ACS applied materials \& interfaces}\ }\textbf
  {\bibinfo {volume} {7}},\ \bibinfo {pages} {25804} (\bibinfo {year}
  {2015})}\BibitemShut {NoStop}%
\bibitem [{\citenamefont {Taillefumier}\ \emph {et~al.}(2002)\citenamefont
  {Taillefumier}, \citenamefont {Cabaret}, \citenamefont {Flank},\ and\
  \citenamefont {Mauri}}]{taillefumier2002}%
  \BibitemOpen
  \bibfield  {author} {\bibinfo {author} {\bibfnamefont {M.}~\bibnamefont
  {Taillefumier}}, \bibinfo {author} {\bibfnamefont {D.}~\bibnamefont
  {Cabaret}}, \bibinfo {author} {\bibfnamefont {A.-M.}\ \bibnamefont {Flank}},
  \ and\ \bibinfo {author} {\bibfnamefont {F.}~\bibnamefont {Mauri}},\
  }\href@noop {} {\bibfield  {journal} {\bibinfo  {journal} {Physical Review
  B}\ }\textbf {\bibinfo {volume} {66}},\ \bibinfo {pages} {195107} (\bibinfo
  {year} {2002})}\BibitemShut {NoStop}%
\end{thebibliography}%

\end{document}